\documentclass[11pt, reqno]{amsart}
\usepackage{amsmath,amssymb}
\usepackage[authoryear,longnamesfirst]{natbib}
\usepackage{graphicx,epsfig,epstopdf}
\usepackage{amssymb}
\usepackage{amsmath}
\usepackage{amsmath}
\usepackage{amsthm}
\usepackage{amsfonts}
\usepackage{graphicx}
\usepackage{bbm}
\usepackage{mathrsfs}
\usepackage{enumerate}
\usepackage{dsfont}
\usepackage{natbib}
\usepackage{multirow}
\usepackage{dsfont}
\usepackage{color}
\newtheorem{theorem}{Theorem}
\newtheorem{assumption}{Assumption}

\newtheorem{corollary}{Corollary}
\newtheorem{lemma}{Lemma}

\theoremstyle{definition}

\newtheorem{remark}{Remark}[section]
\numberwithin{remark}{section}

\newcommand{\Real}{\mathbb R}

\newcommand{\be}{\begin{eqnarray}}
\newcommand{\ee}{\end{eqnarray}}

\newcommand{\NN}{\mathbbm{N}}

\newcommand{\ba}{\begin{array}}
\newcommand{\ea}{\end{array}}
\newcommand{\bs}{\begin{align}\begin{split}\nonumber}
\newcommand{\bsnumber}{\begin{align}\begin{split}}
\newcommand{\es}{\end{split}\end{align}}

\renewcommand{\hat}{\widehat}

\newcommand{\G}{\mathcal{G}}

\newcommand{\GC}{\mathbb{G}_C}

\newcommand{\Ep}{{\mathrm{E}}}

\newcommand{\Op}{O_\Pr}
\newcommand{\op}{o_\Pr}
\renewcommand{\Pr}{{\mathrm{P}}}

\def\supp{{\rm support}}

\newcommand{\semin}[1]{\phi_{{\rm min}}(#1)}
\newcommand{\semax}[1]{\phi_{{\rm max}}(#1)}
\renewcommand{\hat}{\widehat}
\renewcommand{\leq}{\leqslant}
\renewcommand{\geq}{\geqslant}

\newcommand{\I}{\mathcal I}

\newcommand{\argmin}{\rm argmin}

\newcommand{\sumi}{\sum_{i=1}^{N}}
\newcommand{\sumj}{\sum_{j=1}^{M}}
\newcommand{\1}{\mathbbm 1}
\newcommand{\C}{\underline C}
\newcommand{\E}{\mathcal E}
\newcommand{\MM}{\mathcal M}
\newcommand{\PP}{\mathcal P}
\begin{document}

\title[Lasso under multi-way clustering]{Lasso under multi-way clustering:\\
Estimation and Post-selection Inference }

\author[Chiang and Sasaki]{Harold Chiang and Yuya Sasaki}\address[Harold Chiang]{Department of Economics, Vanderbilt University, United States}\address[Yuya Sasaki]{Department of Economics, Vanderbilt University, United States}


\date{First arXiv version:  May 6, 2019.
\\
\indent Code files are available upon request from the authors.}


\maketitle

\begin{abstract}
This paper studies high-dimensional regression models with lasso when data is sampled under multi-way clustering. 
First, we establish convergence rates for the lasso and post-lasso estimators. 
Second, we propose a novel inference method based on a post-double-selection procedure and show its asymptotic validity.
Our procedure can be easily implemented with existing statistical packages. 
Simulation results demonstrate that the proposed procedure works well in finite sample.  
We illustrate the proposed method with a couple of empirical applications to development and growth economics.
\end{abstract}

\section{Introduction}\label{sec:introduction}
This paper studies a method of estimation and post-selection inference for regression parameters in high-dimensional linear models by lasso under multi-way clustering.
The objective is motivated by recently increasing demands from applied economic research.
On one hand, economists often use multi-way cluster sampled data.
Examples include, but are not limited to, network data, matched employer-employee data, matched student-teacher data, scanner data where observations are double-indexed by stores and products, market share data where observations are double-indexed by market and products, and growth/development data where observations are double-indexed by ethnicity and geographical units -- see Section \ref{sec:empirical illustration} for specific applications of the last example.
On the other hand, researchers also often use machine learning methods of estimation and inference for high-dimensional models in today's big data environments.
There are a number of useful methods in the literature that deal with each of these two issues (multi-way clustering and high dimensionality) separately, but the existing methods do not seem to provide a solution to dealing with both of these practically relevant issues simultaneously.
In this light, we present lasso under multi-way clustering, and propose a post-selection inference method for regression parameters under this sampling assumption.

In the important branch of the literature following the seminal work by \cite{BCCH12}, post-selection inference with lasso has been widely studied under various settings by \cite{BCH14}\footnote{See also \cite{BCH11}.}, \cite{JM14}, \cite{vdG14}, \cite{ZZ14}, \cite{BCK15}, and many others. 
For empirical researchers, lasso has become a powerful machine learning tool under data-rich environments. 
Most of the papers in this literature assume i.i.d. or independent sampling. 
In many empirical applications, it is sometimes more plausible to assume multi-way cluster sampling (e.g., network data, matched employer-employee data, and matched student-teacher data).
Building upon \cite{BCCH12}, \cite{BC13} and \cite{BCH14}, this paper generalizes lasso and post-double-selection procedure by allowing for multi-way cluster sampling. 
To our best knowledge, the present paper is the first in the literature of high-dimensional models to consider lasso under multi-way cluster sampling. 

The influential work by \cite{CGM11} proposes multi-way cluster-robust inference methods for linear and nonlinear regression models -- also see \citet[][Section V]{CM15} for a survey. 
Formal analysis of asymptotic properties and bootstrap validity under multi-way clustering is studied by \cite{menzel17} using the Aldous-Hoover representation -- see \citet[][Chapter 7]{kallenberg05} for example. Under the assumptions of separable exchangeability, the method of \cite{menzel17} covers both degenerate and non-degenerate cases.
Using the same representation, while focusing on the non-degenerate cases, \cite{DDG18} develop empirical process theory under multi-way cluster sampling which applies to a large class of econometric models. 
Building upon the asymptotic framework of these two papers, \cite{MNW19} propose several wild bootstrap procedures for linear regression models, and examine their finite-sample performances under several different cluster sampling scenarios.
In this paper, we take advantage of the innovations by these preceding papers to develop a multi-way cluster-robust inference method for high-dimensional models.
To our best knowledge, the present paper is the first in this literature on multi-way clustering to consider high-dimensional models. 

The rest of this paper is organized as follows.
Section \ref{sec:model} introduces the model.
Section \ref{sec:overview} presents an overview of the proposed methodology.
Section \ref{sec:asymptotic_theory} discusses a formal asymptotic theory.
Section \ref{sec:heterogeneous_cluster_sizes} presents an extension of the baseline results to cases of heterogeneous cluster sizes.
Section \ref{sec:simulation_studies} presents simulation studies.
Section \ref{sec:empirical illustration} presents an empirical illustration with development and growth economics.
Section \ref{sec:conclusion} concludes.
The appendix contains mathematical proofs and auxiliary lemmas.

\section{The Model}\label{sec:model}
Consider the high-dimensional regression model
\begin{align}
\label{eq:model_y}
Y_{ij}=&D_{ij}\alpha +X'_{ij}\beta + R^Y_{ij} + \varepsilon_{ij},\:
\Ep[\varepsilon_{ij}|D_{ij},X_{ij}]=0,
\end{align}
where 
$Y_{ij}$ is an observed outcome variable,
$(D_{ij},X_{ij}')'$ is an observed vector of regressors, and
$R_{ij}^Y$ is an approximation error for the unit of observation with the double index $(i,j)$. 
We set $\alpha$ as a scalar parameter of interest.
The dimension $p$ of the nuisance parameter vector $\beta \in \Real^p$ is potentially increasing in the sample size. 
Following the literature on high-dimensional post-selection inference \citep[e.g.,][]{BCH14}, we also consider the auxiliary projection
\begin{align}
\label{eq:model_d}
D_{ij}=&X_{ij}\gamma + R^D_{ij} + v_{ij},\:
\Ep[v_{ij}|X_{ij}]=0,
\end{align}
where $R^D$ is an approximation error. 
The dimension $p$ of the nuisance parameter vector $\gamma\in \Real^p$ is the same as that of $\beta$, and is potentially increasing in the sample size. 

In the absence of two-way clustering, the system (\ref{eq:model_y})--(\ref{eq:model_d}) would be the same as the model considered in \cite{BCH14}. 
We first consider two-way clustering where each cell contains one observation. 
Section \ref{sec:heterogeneous_cluster_sizes} presents an extension to the case of heterogeneous cluster sizes.

\section{Overview of the Method}\label{sec:overview}
In this section, we present an overview of the proposed method, namely estimation and post-selection inference.
Formal theoretical justifications are discussed in Section \ref{sec:asymptotic_theory}.

A researcher observes a sample $\left\{\left. (Y_{ij},D_{ij},X_{ij}') \right\vert i \in \{1,...,N\}, j \in \{1,...,M\}\right\}$ of size $NM$.
The estimation procedure consists of two steps.
First, define the lasso estimates for (\ref{eq:model_y}) and (\ref{eq:model_d}) by
\begin{align}
(\hat \alpha,\hat \beta')'=& \underset{\alpha,\beta }{\argmin} \sumi\sumj (Y_{ij} - \alpha D_{ij} - X_{ij}'\beta)^2 +\lambda_1\|(\alpha,\beta')'\|_1,\label{eq:lasso_main}\\
\text{and}\qquad
\hat \gamma=& \underset{\gamma }{\argmin} \sumi\sumj ( D_{ij} - X_{ij}'\gamma)^2 +\lambda_2 \|\gamma\|_1,\label{eq:lasso_second}
\end{align} 
respectively, for some regularization parameters $\lambda_1$ and $\lambda_2$, valid choices of which are discussed in the statement of Theorem \ref{theorem:rates} ahead.
Denote the supports of the lasso estimates by $\hat I_1=\supp(\hat \beta)$ and $\hat I_2=\supp(\hat \gamma)$, and let $\hat I= \hat I_1 \cup \hat I_2$. 
In the second step, define the post-double-selection lasso estimate $\widetilde \alpha$ by 
\begin{align}
(\widetilde \alpha,\widetilde \beta')= \underset{\supp(\beta) \subset \hat I}{\argmin} \sumi\sumj (Y_{ij} - \alpha D_{ij} - X_{ij}'\beta)^2.\label{eq:post_lasso}
\end{align} 

Let $\C =N\wedge M$, $\mu_N=\C/N$ and $\mu_M=\C/M$. 
Under suitable conditions to be formally stated in Section \ref{sec:asymptotic_theory}, we have the asymptotic normality
\begin{align*}
\sigma^{-1}\sqrt{\C}(\widetilde \alpha - \alpha) \leadsto N(0,1),
\end{align*}
where the asymptotic variance is given by $\sigma^2= Q^{-1}\Gamma Q^{-1}$ with
\begin{align*}
Q=&\Ep[v_{11}^2],
\\
\Gamma=&\bar\mu_N  \Gamma_N + \bar\mu_M  \Gamma_M
=\bar\mu_N\Ep[v_{11} \varepsilon_{11}  v_{12}\varepsilon_{12}  ] + \bar\mu_M \Ep[ v_{11}\varepsilon_{11}v_{21} \varepsilon_{21}],
\end{align*}
and $\bar\mu_N$ and $\bar\mu_M$ denoting the limits of $\mu_N$ and $\mu_M$, respectively.

The asymptotic variance is estimated by the sample counterpart $\hat\sigma^2= \hat Q^{-1}\hat\Gamma \hat Q^{-1}$, where
\begin{align*}
\hat Q=&\frac{1}{NM}\sumi\sumj \hat v_{ij}^2,
\\
\hat \Gamma
=& \frac{\C}{(NM)^2}\sumi\sum_{1\le j,j'\le M} \hat v_{ij} \hat \varepsilon_{ij} \hat  \varepsilon_{ij'}  \hat  v_{ij'}
+ \frac{\C}{(NM)^2}\sum_{1\le i,i'\le N}\sumj \hat v_{ij}\hat \varepsilon_{ij} \hat\varepsilon_{i'j}\hat v_{i'j} ,
\end{align*}
$\hat v_{ij}= D_{ij}- X_{ij}'\hat \gamma$, and $\hat \varepsilon_{ij}=Y_{ij}- \hat \alpha D_{ij} - X_{ij}'\hat\beta$.

In summary, we propose to report the post-double-selection lasso estimate $\widetilde\alpha$ as an estimate of $\alpha$ with its standard error given by $\hat\sigma / \sqrt{\C}$.
The $\alpha^\ast$-level confidence interval can be constructed as $\left[\widetilde\alpha + \Phi^{-1}(\alpha^\ast/2)\hat\sigma / \sqrt{\C},\widetilde\alpha + \Phi^{-1}(1-\alpha^\ast/2)\hat\sigma / \sqrt{\C}\right]$, where $\Phi^{-1}$ denotes the quantile function of the standard normal distribution.

\section{Asymptotic Theory}\label{sec:asymptotic_theory}

The two-way sample sizes $(N,M) \in \mathbb{N}^2$ will be index by a single index $n \in \mathbb{N}$ as $(N,M) = (N(n),M(n))$ where $M(n)$ and $N(n)$ are non-decreasing in $n$ and $M(n)N(n)$ is increasing in $n$. For simplicity, each size of intersection $n_{ij}$ is assumed to be uniformly bounded by a positive integer $\bar n$ that is independent of $n$.
With this said, we will suppress the index notation and write $(N,M)$ for simplicity.
We fix a number of notations.
For each $n$, let $\Pr_{n}$ denote the law with respect to sample size $(N,M)$ -- note that we allow the dimension $p$ of $X_{ij}$ to grow with $n$.
Let $a := p\vee (NM)$.
Also recall the notations $\C =N\wedge M$, $\mu_N=\C/N$, and $\mu_M=\C/M$ from Section \ref{sec:overview}.
We use the short-hand notation $[k] = \{1,...,k\}$ and $[k]^c=\mathbb{N} \backslash [k]$ for any $k \in \mathbb{N}$.
For a sequence $(t_{ij})_{i\in[N],j\in [M]}$, denote $\|t_{ij}\|_n = \sqrt{\frac{1}{NM} \sumi\sumj t_{ij}^2 }$. 
Thus, $\|X'_{ij}\delta\|_n = \sqrt{\frac{1}{NM} \sumi\sumj \delta' X_{ij}X_{ij}'\delta}$ is the prediction norm of $\delta$.
Let $\|A\|_\infty=\max_{k,l}|A_{k,l}|$ denote the max norm of matrix $A$.
We write $a \lesssim b$ to mean $a \leq cb$ for some $c > 0$ that does not depend on $n$.
We also write $a \lesssim_\Pr b$ to mean $a = O_P(b)$.
We write $Z_{ij}=(Y_{ij},D_{ij},X'_{ij})'$ for the $(p+2)$-dimensional random vector in data.
Throughout, we assume that this random vector $Z_{ij}$ is Borel measurable -- see \citet[][pp. 304]{kallenberg05}.
With these notations, we state the following four assumptions.

\begin{assumption}[Sampling]\label{a:sampling}
Suppose that $\C \to \infty $, $\mu_N \to \bar \mu_N \geq 0$, and $\mu_M \to \bar \mu_M \geq 0$.
\begin{enumerate}
\item $(Z_{ij})_{(i,j)\in \mathbbm N^2}$ is an infinite sequence of separately exchangeable $(p+2)$-dimensional random vectors. 
That is, for any permutations $\pi_1$ and $\pi_2$ of $\mathbbm N$, we have
\begin{align*}
(Z_{ij})_{(i,j)\in \mathbbm N^2}\overset{d}{=} (Z_{\pi_1(i)\pi_2(j)})_{(i,j)\in \mathbbm N^2}.
\end{align*}
\item $(Z_{ij})_{(i,j)\in \mathbbm N^2}$ is dissociated. 
That is, for any $(c_1,c_2)\in \mathbbm N^2$, 
$
(Z_{ij})_{i \in [c_1], j \in [c_2]}
$
is independent of 
$
(Z_{ij})_{i \in [c_1]^c, j \in [c_2]^c}.
$
\item For each $n$, an econometrician observes $(Z_{ij})_{i\in[N],j\in[M]}$.
\end{enumerate}
\end{assumption}

\begin{assumption}[Moments]\label{a:moments}
There exists a sequence $\{B_n\}_{n=1}^\infty$ of positive constants such that the following conditions hold for all $n \in \mathbb{N}$ for some $q>4$:
\begin{enumerate}
\item 
$\Ep[|D_{11}|^{2q}]+\max_{k\in [p]}\Ep[|X_{11,k}|^{2q}]+ \Ep[|\varepsilon_{11}|^{2q}|X_{11},v_{11}]+\Ep[|v_{11}|^{2q}|X_{11}]\le K$ a.s. 
and 
$0 < c\le \Ep[v^{2}_{11}|X_{11}]$ a.s. for positive constants, $c$ and $K$, that are independent of $n$.
\item $\Ep[\|X_{11}\|_\infty^{2q}]\le B_{n}^{2q}$ and $B_n\sqrt{\log a}\lesssim (N\vee M)^{1/2-1/q}$.
\item $\bar \mu_N\Ep[v_{11} \varepsilon_{11}  v_{12}\varepsilon_{12}  ] + \bar \mu_M \Ep[ v_{11}\varepsilon_{11}v_{21} \varepsilon_{21}] $ and the maximal and minimal eigenvalues of $\Ep[X_{11}X_{11}']$ are bounded and bounded away from zero uniformly in $n$.
\end{enumerate}
\end{assumption}

\begin{assumption}[Sparsity]\label{a:sparsity}
${}$
\begin{enumerate}
\item $\|\beta\|_0+ \|\gamma\|_0 \lesssim s$.
\item $\|R_{ij}^Y\|_n+ \|R_{ij}^D\|_n\le c_s \lesssim_\Pr \sqrt{s /\C}$.
\item $\frac{s^2_n(\log (a))^2}{\C}=o(1)$.
\end{enumerate}
\end{assumption}

\begin{assumption}[Sparse Eigenvalues]\label{a:sparse_eigenvalues}
There exists a sequence $\{\ell_n\}$ such that $\ell_n \to \infty$ and, with probability at least $1-o(1)$,
\begin{align*}
0<c\le \semin{s\ell_n}\le \semax{s\ell_n}\le c' <\infty
\end{align*}
holds for some constants, $c$ and $c'$, that are independent of $n$, where
\begin{align*}
\semax{m}:=
\max_{\substack{1\le \|\delta\|_0 \le m }} \frac{\delta' M\delta}{\|\delta\|^2}
\text{ and } \semin{m}:= \min_{\substack{1\le \|\delta\|_0 \le m }} \frac{\delta' M\delta}{\|\delta\|^2},
\end{align*}
with $M=\frac{1}{NM} \sumi\sumj X_{ij}X_{ij}'$, denote the maximal and minimal $m$-sparse eigenvalues.
\end{assumption}

\begin{remark}[Discussion of the Assumptions]\label{remark:discussion_assumptions}
Assumption \ref{a:sampling} is closely related to Assumption 1 of \cite{DDG18}. 
The main difference is that we allow $p$ to be changing with $n$. 
We remark that the exchangeability assumption is not new in econometrics -- it has been used in \cite{andrews05} and \cite{menzel15} as well as \citet{menzel17}, \citet{DDG18}, and \citet{MNW19}.
Assumption \ref{a:moments} is standard in the literature on post-selection inference with lasso. 
Parts (1) and (2) require an existence of higher order moments of key objects. 
Note that common assumptions in high-dimensional literature, such as sub-gaussianity or boundedness, are not required. 
They can be replaced by some higher level conditions similar to Condition RF of \cite{BCCH12}.\footnote{See their Lemma 3.} 
Part (3) of Assumption \ref{a:moments} requires that the asymptotic variance is bounded away from zero.\footnote{Similarly to \cite{DDG18}, we focus on non-degenerate cases in this paper. See \cite{menzel17} for the studies of degenerate cases using a bootstrap-based method.}
Assumption \ref{a:sparsity} is a direct generalization of Condition ASTE (iii) and (iv) of \cite{BCH14}. 
Finally, Assumption \ref{a:sparse_eigenvalues} is analogous to Condition SE of \cite{BCH14}, which is standard in the high-dimensional literature. 
It only imposes small diagonal submatrices to be well behaved.
$\triangle$
\end{remark}

\subsection{Independentization via H\'ajek Projection}
In this section, we show that an empirical process in multi-way clustered samples can be represented as a sum of independent variables via H\'ajek projection.
Furthermore, its variance can be shown to be approximated by covariances of observed variables.

For any $f:\supp(Z)\to  \Real$, we let
\begin{align*}
\GC f:=\sqrt{\C}\Big\{\frac{1}{NM} \sumi\sumj f(Z_{ij}) - \Ep[f(Z_{11})]\Big\}
\end{align*}
denote its empirical process.

\begin{lemma}[Independentization via H\'ajek Projection]\label{lemma:hajek}
If Assumption \ref{a:sampling} holds and $f:\supp(Z)\to  \Real$ satisfies $\Ep f^2(Z_{11})<K$ for a finite constant $K$ that is independent of $n$, then there exist i.i.d. uniform random variables $U_{i0}$ and $U_{0j}$ such that the H\'ajek projection $H_n f$ of $\GC f$ on 
$$
\G_n=\Big\{ \sumi g_{i0}(U_{i0}) + \sumj g_{0j}(U_{0j}) : g_{i0}, g_{0j} \in L^2(\Pr_{n}) \Big\}
$$
is equal to
\begin{align*}
H_n f=\sumi \frac{\sqrt{\C}}{N}  \Ep\Big[f(Z_{i1})- \Ep f(Z_{11}) \Big| U_{i0}\Big] + \sumj\frac{\sqrt{\C}}{M} \Ep\Big[f(Z_{1j})- \Ep f(Z_{11}) \Big| U_{0j}\Big]
\end{align*}
for each $n$.
Furthermore,
\begin{align*}
V(\GC f)= V(H_n f)+O(\C^{-1})=\bar\mu_N  Cov(f(Z_{11}),f(Z_{12})) + \bar\mu_M Cov(f(Z_{11}),f(Z_{21}))+O(\C^{-1})
\end{align*}
holds a.s.
\end{lemma}

A proof of this lemma can be found in Appendix \ref{sec:lemma:hajek}.
The first part of the lemma shows that an empirical process $\GC f$ under multi-way cluster sampling can be represented as a sum of independent unobserved variables via H\'ajek projection $H_n f$.
While $U_{i0}$ and $U_{0j}$ are unobserved, the second part of this lemma in turn shows that the variance of the H\'ajek projection can be approximated by covariances of observed variables.
Note that, since $H_n f$ is a H\'ajek projection, the lemma implies $\frac{\GC f}{\sqrt{V(\GC f)}}=\frac{H_n f}{\sqrt{V(H_n f)}}+\op(1)$ if $\bar\mu_N  Cov(f(Z_{11}),f(Z_{12})) + \bar\mu_M Cov(f(Z_{11}),f(Z_{21}))$ is bounded and bounded away from zero uniformly in $n$.

Our Lemma \ref{lemma:hajek} can be seen as an extension to Lemma D.2 in \cite{DDG18}.
Specifically, while \cite{DDG18} consider a fixed data generating process over the sample size $n$, our Lemma \ref{lemma:hajek} allows the data generating process to vary with $n$ in particular for the sake of accommodating the increasing of dimensionality $p$ for high-dimensional models.
The lemma serves as a main building block for all the asymptotic results to be presented ahead.

\subsection{Convergence Rates of Lasso and Post-Lasso under Multi-Way Clustering}

We next show the convergence rates of the lasso estimator $(\widehat\alpha,\widehat\beta',\widehat\gamma')'$ and the post-lasso estimator $(\widetilde\alpha,\widetilde\beta',\widetilde\gamma')'$ under multi-way clustering.

\begin{theorem}[Convergence Rates for Lasso and Post-Lasso under Multi-Way Clustering]\label{theorem:rates}
If Assumptions \ref{a:sampling}, \ref{a:moments} (1)--(2), \ref{a:sparsity} (1)--(2), and \ref{a:sparse_eigenvalues} are satisfied, and $\lambda_1, \lambda_2 =C\sqrt{(NM)^2 \log a/\C}$ for some constant $C>1$, then
\begin{align*}
&\|\hat\eta-\eta\|_1 + \|\hat \gamma - \gamma\|_1 \lesssim \sqrt{\frac{s^2\log a}{\C}},\, \qquad  \|W_{ij}'(\hat\eta-\eta)\|_n + \|X_{ij}(\hat \gamma - \gamma)\|_n\lesssim \sqrt{\frac{s\log a}{\C}},\\
&\|\widetilde\eta-\eta\|_1 + \|\widetilde \gamma - \gamma\|_1 \lesssim \sqrt{\frac{s^2\log a}{\C}},\, \qquad \|W_{ij}'(\widetilde\eta-\eta)\|_n + \|X_{ij}(\widetilde \gamma - \gamma)\|_n\lesssim \sqrt{\frac{s\log a}{\C}}, \quad\text{and}\\
&\|\hat\eta-\eta\| + \|\hat \gamma - \gamma\| + \|\widetilde\eta-\eta\| + \|\widetilde \gamma - \gamma\| \lesssim \sqrt{\frac{s\log a}{\C}}
\end{align*}
hold, where $W_{ij}=[D_{ij},X_{ij}']'$ and $\eta=(\alpha,\beta')'$.
\end{theorem}

A proof can be found in Appendix \ref{sec:theorem:rates}, and is based on the previous result (Lemma \ref{lemma:hajek}).
In the multi-way sampling, this lemma can be viewed as a counterpart of Lemma 6 and Lemma 7 in \cite{BCCH12}.

\subsection{Post-Selection-Inference with Post-Lasso under Multi-way Clustering}\label{sec:post_selection_inference}

In this section, we present the main result of this paper.
The limit normal distribution of the post-double-selection lasso estimate $\widetilde\alpha$ is established based on the previous two results (Lemma \ref{lemma:hajek} and Theorem \ref{theorem:rates}). 

\begin{theorem}[Asymptotic Normality]\label{theorem:asymptotic_normality}
If Assumptions \ref{a:sampling}, \ref{a:moments}, \ref{a:sparsity} and \ref{a:sparse_eigenvalues} are satisfied, and $\lambda_1$ and $\lambda_2$ are chosen according to the statement of Theorem \ref{theorem:rates}, then
\begin{align*}
\sigma^{-1}\sqrt{\C}(\widetilde \alpha - \alpha) \leadsto N(0,1),
\end{align*}
where $\sigma^2= Q^{-1}\Gamma Q^{-1}$, $Q=\Ep[v_{11}^2]$ and 
\begin{align*}
\Gamma=&\bar\mu_N  \Gamma_N + \bar\mu_M  \Gamma_M
=\bar\mu_N\Ep[v_{11} \varepsilon_{11}  v_{12}\varepsilon_{12}  ] + \bar\mu_M \Ep[ v_{11}\varepsilon_{11}v_{21} \varepsilon_{21}].
\end{align*}
\end{theorem}

A proof can be found in Appendix \ref{sec:theorem:asymptotic_normality}.
This result provides a theoretical justification for the asymptotic variance proposed in the overview in Section \ref{sec:overview}.
In practice, we do not know the components, $Q$ and $\Gamma$, of the asymptotic variance.
The following subsection proposes estimators of them.

\subsection{Variance Estimation}

In this section, we propose an analog variance estimator.
The components, $Q$ and $\Gamma$, of the asymptotic variance can be estimated by
\begin{align*}
\hat Q=&\frac{1}{NM}\sumi\sumj \hat v_{ij}^2
\qquad\text{and}
\\
\hat \Gamma
=& \frac{\C}{(NM)^2}\sumi\sum_{1\le j,j'\le M} \hat v_{ij} \hat \varepsilon_{ij} \hat  \varepsilon_{ij'}  \hat  v_{ij'}
+ \frac{\C}{(NM)^2}\sum_{1\le i,i'\le N}\sumj \hat v_{ij}\hat \varepsilon_{ij} \hat\varepsilon_{i'j}\hat v_{i'j} ,
\end{align*}
respectively, where $\hat v_{ij}= D_{ij}- X_{ij}'\hat \gamma$ and $\hat \varepsilon_{ij}=Y_{ij}- \hat \alpha D_{ij} - X_{ij}'\hat\beta$ are the residuals.
With these component estimators, we propose that the asymptotic variance $\sigma^2 = Q^{-1} \Gamma Q^{-1}$ be estimated by $\hat\sigma^2= \hat Q^{-1}\hat\Gamma \hat Q^{-1}$.
The following theorem provides a theoretical support for this variance estimator.

\begin{theorem}[Variance Estimation]\label{theorem:variance_est}
If Assumptions \ref{a:sampling}, \ref{a:moments}, \ref{a:sparsity} and \ref{a:sparse_eigenvalues} are satisfied, $\lambda_1$ and $\lambda_2$ are chosen according to the statement of Theorem \ref{theorem:rates},
$\frac{(NM)^{1/q} B_n^2 s^3 (\log a)^2}{\C ^2}=o(1)$,
$\frac{(NM)^{1/q}  s \log a}{ \C}=o(1)$, and
$\|R^D_{ij}R^Y_{ij}\|^2_n=O(1)$,
 then the variance estimator $\hat\sigma^2= \hat Q^{-1}\hat\Gamma \hat Q^{-1}$ is consistent for $\sigma^2 = Q^{-1} \Gamma Q^{-1}$.
\end{theorem}

A proof is found in Appendix \ref{sec:theorem:variance_est}.
In light of this result, we propose to compute the standard error by $\hat\sigma / \sqrt{\C}$.
Similarly, in light of this result together with Theorem \ref{theorem:asymptotic_normality}, we propose to construct the $\alpha^\ast$-level confidence interval by $\left[\widetilde\alpha + \Phi^{-1}(\alpha^\ast/2)\hat\sigma / \sqrt{\C},\widetilde\alpha + \Phi^{-1}(1-\alpha^\ast/2)\hat\sigma / \sqrt{\C}\right]$, where $\Phi^{-1}$ denotes the quantile function of the standard normal distribution.

\section{Extension: Heterogeneous Cluster Sizes}\label{sec:heterogeneous_cluster_sizes}
Thus far, we focus on the case where each cluster contains one observation.
In this section, we presented an extension of the baseline results to situations where the numbers of observations are heterogeneous across clusters. 
Suppose that we have $n_{ij}$ observations for each cell $(i,j) \in [N] \times [M]$, where $n_{ij}$ is a random variable that is allowed to depend on $(X_{ij,\ell})_{\ell\ge 1}$. To deal with the situation of $n_{ij}=0$, for any sequence $(t_\ell)_{\ell \ge 1}$, define $\sum_{\ell=1}^0 t_\ell=0$.
Consider the model
\begin{align*}
Y_{ij,\ell}=&D_{ij,\ell}\alpha +X_{ij,\ell}'\beta + R^Y_{ij,\ell} + \varepsilon_{ij,\ell},\:
\Ep[\varepsilon_{ij,\ell}|D_{ij,\ell},X_{ij,\ell}]=0,
\end{align*}
where 
$Y_{ij,\ell}$ is an observed outcome variable,
$(D_{ij,\ell},X_{ij,\ell}')'$ is an observed vector of regressors, and
$R_{ij,\ell}^Y$ is an approximation error for the unit $\ell\in [n_{ij}]$ with the double index $(i,j)$ indicating $i$-th cluster in the first clustering dimension and $j$-th cluster in the second clustering dimension. 
Using matrix notations, we can rewrite the model as
\begin{align*}
Y_{ij}=&D_{ij}\alpha +X_{ij}\beta + R^Y_{ij} + \varepsilon_{ij},\:
\Ep[\varepsilon_{ij}|D_{ij},X_{ij}]=0,
\end{align*}
where each of 
$Y_{ij}=(Y_{ij,\ell})_{\ell\in [n_{ij}]}$, 
$D_{ij}=(D_{ij,\ell})_{\ell\in [n_{ij}]}$, 
$R^Y_{ij}=(R^Y_{ij,\ell})_{\ell\in [n_{ij}]}$, 
and
$\varepsilon_{ij}=(\varepsilon_{ij,\ell})_{\ell\in [n_{ij}]}$ is of dimension $n_{ij}\times 1$,
and 
$X_{ij}=(X_{ij,\ell}')_{\ell\in [n_{ij}]}$ is of dimension $n_{ij}\times p$.
We similarly write the accompanying auxiliary projection as
\begin{align*}
D_{ij}=&X_{ij}\gamma + R^D_{ij} + v_{ij},\:
\Ep[v_{ij}|X_{ij}]=0,
\end{align*}
where $R^D$ is of dimension $n_{ij}\times 1$ representing approximation errors, and $v_{ij}$ is of dimension $n_{ij}\times 1$ representing projection errors. 

Under this setting, the first step of estimation procedure consists of
\begin{align}
(\hat \alpha,\hat \beta')'=& \underset{\alpha,\beta }{\argmin} \sumi\sumj\sum_{\ell=1}^{n_{ij}} (Y_{ij,\ell} - \alpha D_{ij,\ell} - X_{ij,\ell}'\beta)^2 +\lambda_1\|(\alpha,\beta')'\|_1\nonumber\\
\text{and}\qquad
\hat \gamma=& \underset{\gamma }{\argmin} \sumi\sumj\sum_{\ell=1}^{n_{ij}} ( D_{ij,\ell} - X_{ij,\ell}'\gamma)^2 +\lambda_2 \|\gamma\|_1.\nonumber
\end{align} 
In turn, the second-step estimates are obtained by
\begin{align}
(\widetilde \alpha,\widetilde \beta')= \underset{\supp(\beta) \subset \hat I}{\argmin} \sumi\sumj\sum_{\ell=1}^{n_{ij}}  (Y_{ij,\ell} - \alpha D_{ij,\ell} - X_{ij,\ell}'\beta)^2.\nonumber
\end{align} 
The asymptotic variance estimator for $\widetilde\alpha$ is given by $\hat\sigma^2= \hat Q^{-1}\hat\Gamma \hat Q^{-1}$, where
\begin{align*}
\hat Q=&\frac{1}{NM}\sumi\sumj \hat v_{ij}'\hat v_{ij},
\\
\hat \Gamma
=& \frac{\C}{(NM)^2}\sumi\sum_{1\le j,j'\le M} \hat v_{ij}' \hat \varepsilon_{ij}\hat  \varepsilon_{ij'}'  \hat  v_{ij'}
+ \frac{\C}{(NM)^2}\sum_{1\le i,i'\le N}\sumj \hat v_{ij}'\hat \varepsilon_{ij}  \hat\varepsilon_{i'j}' \hat v_{i'j},
\end{align*}
$\hat v_{ij}= D_{ij}- X_{ij}\hat \gamma$, and $\hat \varepsilon_{ij}=Y_{ij}- \hat \alpha D_{ij} - X_{ij}\hat\beta$.

We now formally state assumptions for the extended theory to support the asymptotic validity of this procedure.
Define
$W_{ij}=(n_{ij},(Z_{ij,\ell})_{\ell \ge 1})$ and $\ddot M=\frac{1}{NM}\sumi\sumj X_{ij}'X_{ij}$.
\begin{assumption}[Sampling]\label{a:sampling_hetero}
Suppose that $\C \to \infty $, $\mu_N \to \bar \mu_N \geq 0$, and $\mu_M \to \bar \mu_M \geq 0$.
\begin{enumerate}
\item $(W_{ij})_{(i,j)\in \mathbbm N^2}$ is an infinite sequence of separately exchangeable random processes.
\item $(W_{ij})_{(i,j)\in \mathbbm N^2}$ is dissociated. 
\item For each $n$, an econometrician observes $((W_{ij,\ell})_{\ell\in [n_{ij}]} )_{i\in[N],j\in[M]}$.
\item $\Ep[n_{ij}]>0$ and $n_{ij}\le \bar n$ for a positive finite constant $\bar n$ independent of $n$.
\end{enumerate}
\end{assumption}

\begin{assumption}[Moments]\label{a:moments_hetero}
There exists a sequence $\{B_n\}_{n=1}^\infty$ of positive constants such that the following conditions hold for all $n \in \mathbb{N}$ for some $q>4$:
\begin{enumerate}
\item 
$\Ep[\max_{\ell\in [n_{ij}]}|D_{11,\ell}|^{2q}]+\max_{k\in [p]}\Ep[\max_{\ell\in [n_{ij}]}|X_{11,\ell,k}|^{2q}]+ \Ep[\max_{\ell\in [n_{ij}]}|\varepsilon_{11,\ell}|^{2q}|X_{11,\ell},v_{11,\ell}]+\Ep[|v_{11,\ell}|^{2q}|X_{11,\ell}]\le K$ a.s. 
and 
$0 < c\le \Ep[\max_{\ell\in [n_{ij}]}v^{2}_{11,\ell}|X_{11,\ell}]$ a.s. for positive constants, $c$ and $K$, that are independent of $n$.
\item $\Ep[\max_{\ell\in [n_{ij}]}\|X_{11}\|_\infty^{2q}]\le B_{n}^{2q}$ and $B_n\sqrt{\log a}\lesssim (N\vee M)^{1/2-1/q}$.
\item $\bar \mu_N\Ep[v_{11}' \varepsilon_{11}  \varepsilon_{12}'v_{12}  ] + \bar \mu_M \Ep[ v_{11}'\varepsilon_{11}\varepsilon_{21}'v_{21} ] $ and the maximal and minimal eigenvalues of $\Ep[X_{11}'X_{11}]$ are bounded and bounded away from zero uniformly in $n$.
\end{enumerate}
\end{assumption}

\begin{assumption}[Sparsity]\label{a:sparsity_hetero}
${}$
\begin{enumerate}
\item $\|\beta\|_0+ \|\gamma\|_0 \lesssim s$.
\item $\sqrt{(NM)^{-1}\sumi\sumj\sum_{\ell \in [n_{ij}]}(R_{ij,\ell}^Y)^2}+ \sqrt{(NM)^{-1}\sumi\sumj\sum_{\ell \in [n_{ij}]}(R_{ij,\ell}^D)^2}\le c_s \lesssim_\Pr \sqrt{s /\C}$.
\item $\frac{s^2_n(\log (a))^2}{\C}=o(1)$.
\end{enumerate}
\end{assumption}

\begin{assumption}[Sparse Eigenvalues]\label{a:sparse_eigenvalues_hetero}
There exists a sequence $\{\ell_n\}$ such that $\ell_n \to \infty$ and, with probability at least $1-o(1)$,
\begin{align*}
0<c\le \semin{s\ell_n}\le \semax{s\ell_n}\le c' <\infty
\end{align*}
holds for some constants, $c$ and $c'$, that are independent of $n$, where
\begin{align*}
\semax{m}:=
\max_{\substack{1\le \|\delta\|_0 \le m }} \frac{\delta' \ddot M\delta}{\|\delta\|^2}
\text{ and } \semin{m}:= \min_{\substack{1\le \|\delta\|_0 \le m }} \frac{\delta'\ddot M\delta}{\|\delta\|^2}.
\end{align*}
\end{assumption}

The following statement provides a theoretical guarantee for the estimation and inference procedure for the extended model outlined above. 
\begin{corollary}\label{corollary:hetero}
If Assumptions \ref{a:sampling_hetero}, \ref{a:moments_hetero}, \ref{a:sparsity_hetero} and \ref{a:sparse_eigenvalues_hetero} are satisfied, and $\lambda_1$ and $\lambda_2$ are chosen according to the statement of Theorem \ref{theorem:rates}, then
\begin{align*}
\sigma^{-1}\sqrt{\C}(\widetilde \alpha - \alpha) \leadsto N(0,1),
\end{align*}
where $\sigma^2= Q^{-1}\Gamma Q^{-1}$ with
\begin{align*}
Q=&\Ep[v_{11}'v_{11}],
\\
\Gamma
=&\bar\mu_N\Ep[v_{11}' \varepsilon_{11}  v_{12}'\varepsilon_{12}  ] + \bar\mu_M \Ep[ v_{11}'\varepsilon_{11}v_{21}' \varepsilon_{21}].
\end{align*}
Furthermore, if $\frac{(NM)^{1/q} B_n^2 s^3 (\log a)^2}{\C ^2}=o(1)$,
$\frac{(NM)^{1/q}  s \log a}{ \C}=o(1)$, and
$\|R^D_{ij}R^Y_{ij}\|^2_n=O(1)$,
then the variance estimator $\hat\sigma^2$ is consistent for $\sigma^2$.
\end{corollary}
A proof of Corollary \ref{corollary:hetero} closely follows that of the results in Section \ref{sec:asymptotic_theory}, and are therefore omitted. 
The key difference is that we now apply Aldous-Hoover representation on $W_{ij}$ rather than on $Z_{ij}$.\footnote{For more insights on this extension, see Section 3.1 of \cite{DDG19}.}

\section{Simulation Studies}\label{sec:simulation_studies}
In this section, we present simulation studies of finite-sample performance of the proposed method of estimation and post-selection inference.
We compare the performance of our method against existing alternatives from the lasso literature that do not account for multi-way clustering.

\subsection{Simulation Setup}
We consider the linear model
\begin{align*}
Y_{ij} = D_{ij} \alpha + X_{ij}' \beta + \varepsilon_{ij}.
\end{align*}
The parameter values are fixed at $(\alpha,\beta')' = \left(0.5, 0.5^2, \cdots, 0.5^{\text{dim}(X)+1}\right)'$.
The random vector $(D_{ij}, X_{ij}', \varepsilon_{ij})$ is constructed by
\begin{align*}
\left(D_{ij}, X_{ij}\right)
=&
(1-\omega^{x}_1-\omega^{x}_2) \upsilon^{x}_{ij} + \omega^{x}_1 \upsilon^{x}_{i} + \omega^{x}_2 \upsilon^{x}_{j}
\qquad\text{and}\\
\varepsilon_{ij}
=&
(1-\omega^{\varepsilon}_1-\omega^{\varepsilon}_2) \upsilon^{\varepsilon}_{ij} + \omega^{\varepsilon}_1 \upsilon^{\varepsilon}_{i} + \omega^{\varepsilon}_2 \upsilon^{\varepsilon}_{j}
\end{align*}
with two-way clustering weights $(\omega^{x}_1, \omega^{x}_2)$ and $(\omega^{\varepsilon}_1,\omega^{\varepsilon}_2)$,
where $\upsilon^{x}_{ij}$, $\upsilon^{x}_{i}$, and $\upsilon^{x}_{j}$ are independently generated according to
\begin{align*}
\upsilon^{x}_{ij}, \upsilon^{x}_{i}, \upsilon^{x}_{j} \sim N\left(0,\left(\begin{array}{ccccc}\rho^0 & \rho^1 & \cdots & \rho^{\text{dim}(X)-1} & \rho^{\text{dim}(X)} \\ \rho^1 & \rho^0 & \cdots & \rho^{\text{dim}(X)-2} & \rho^{\text{dim}(X)-1} \\ \vdots & \vdots & \ddots & \vdots & \vdots \\ \rho^{\text{dim}(X)-1} & \rho^{\text{dim}(X)-2} & \cdots & \rho^0 & \rho^1 \\ \rho^{\text{dim}(X)} & \rho^{\text{dim}(X)-1} & \cdots & \rho^1 & \rho^0 \end{array}\right)\right),
\end{align*}
and $\upsilon^{\varepsilon}_{ij}$, $\upsilon^{\varepsilon}_{i}$, and $\upsilon^{\varepsilon}_{j}$ are independently generated according to
\begin{align*}
\upsilon^{\varepsilon}_{ij}, \upsilon^{\varepsilon}_{i}, \upsilon^{\varepsilon}_{j} \sim N(0,1).
\end{align*}
Note that the weights $(\omega^{x}_1, \omega^{x}_2)$ and $(\omega^{\varepsilon}_1,\omega^{\varepsilon}_2)$ specify the extent of dependence in two-way clustering in $(D_{ij},X_{ij}')$ and $\varepsilon_{ij}$, respectively.
Also, the parameter $\rho$ specifies the extent of collinearity among the high-dimensional covariates $(D_{ij},X_{ij}')$.
We set
$(\omega^{x}_1, \omega^{x}_2) = (0.25,0.25)$,
$(\omega^{\varepsilon}_1,\omega^{\varepsilon}_2) = (0.25,0.25)$, and
$\rho = 0.50$.

\subsection{Alternative Variance Estimators}
We compare the performance of our multi-way cluster-robust variance estimator with two existing alternative benchmarks. 
One is the heteroskedasticity robust variance estimator (such as the one in \cite{BCH14}) without accounting for cluster sampling, i.e., $\Gamma$ is estimated by
\begin{align*}
\hat \Gamma_{HC}=\frac{1}{NM} \sumi\sumj \hat v_{ij}^2 \hat \varepsilon_{ij}^2.
\end{align*}
We will refer to this variance estimator $\widehat Q^{-1} \hat \Gamma_{HC} \widehat Q^{-1}$ as the `0-Way' estimator.
The other is the one-way cluster-robust variance estimator (similar to those of \cite{BCHK16} and \cite{kock16}) clustered at one (e.g., second) dimension, i.e., $\Gamma$ is estimated by
\begin{align*}
\hat \Gamma_{CR}=\frac{1}{NM^2} \sumi\sum_{1\le j,j'\le M} \hat v_{ij}\hat \varepsilon_{ij}\hat v_{ij'}\hat \varepsilon_{ij'}.
\end{align*}
We will refer to this variance estimator $\widehat Q^{-1} \hat \Gamma_{CR} \widehat Q^{-1}$ as the `1-Way' estimator.

\subsection{Results}

Table \ref{tab:simulation_results} summarizes simulation results. 
The first two columns indicate the two-way sample sizes $(N,M)$.
The third column indicates the dimension (Dim) of $(\alpha,\beta')'$. 
The next four columns report simulation statistics for $\widetilde\alpha$.
These statistics include the average (Avg), bias (Bias), standard deviation (SD), and root mean square error (RMSE). 
The last three columns report 95\% coverage frequencies of $\alpha$ based on three variance estimators.
The first is the heteroskedasticity robust variance estimator (0-Way).
The second is the one-way cluster-robust variance estimator (1-Way).
The third is our multi-way cluster-robust variance estimator (2-Way).
The results are based on 25,000 Monte Carlo iterations for each row in the table.

In view of the statistics columns, observe that the post-double-selection lasso estimate $\widetilde\alpha$ behaves well in larger sample sizes (e.g., $N, M \geq 20$) both in terms of bias and variance.
Next, observe the 95\% coverage frequencies by the three alternative variance estimators.
Both the 0-Way and 1-Way variance estimators significantly underestimate the variances of the post-double-selection lasso estimate $\widetilde\alpha$.
On the other hand, the coverage frequency based on our 2-Way variance estimator approaches the nominal probability (95\%) as the sample size increases.
These results demonstrate that, when the true sampling process entails multi-way clustering, traditional variance estimators may bias the inference and our multi-way cluster-robust variance estimator performs robustly well.

\section{Empirical Illustrations}\label{sec:empirical illustration}

In this section, we illustrate our proposed method with applications to a couple of empirical studies.
There is a sequence of recent growth and development economic studies using empirical data that are clustered at ethnic and geographical levels \citep[e.g.,][]{NunnWantchekon11,Michalopoulos_Papaioannou2013,MP14,MP16,gershman16,anderson18,dickens18}.
The next two subsections present how our method can enrich the model flexibility and robustness of such studies, focusing on the cases of \citet{NunnWantchekon11} and \citet{Michalopoulos_Papaioannou2013}.

\subsection{Slave Trade and Mistrust in Africa}\label{sec:slave trade and mistrust in africa}
\citet{NunnWantchekon11} analyze the effects of slave trade on mistrust in Africa, controlling for various demongraphic and geographical covariates including age, age squared, ethnic fractionalization, gender, urban residence, occupation, religion, and living conditions as well as country fixed effects in their baseline model.
Estimates of these effects are obtained by running regressions with a sample that pools $n_{ij}$ individuals $\ell \in [n_{ij}]$ in ethnic group $i$ and districts $j$ across the cells $(i,j) \in [N] \times [M]$ of $N (=185)$ ethnic groups and $M (=1257)$ districts.
Standard errors are computed by the two-way cluster-robust method of \citet{CGM11} for the ethnic group and district as two ways of clustering.

With our proposed method that is applicable to both high-dimensional models and multi-way clustering, they could consider even more flexible model specifications, for example, allowing for higher orders of age rather than just the quadratic specification and interactions of the age polynomials with various other dummy variables.
We present estimates with standard errors under such extended models with flexible specifications, demonstrate that qualitatively similar results continue to be obtained without substantial loss of statistical significance, and thus confirm further robustness of the main empirical findings by \citet{NunnWantchekon11}.

Consider the model
\begin{align*}
Y_{ij,\ell}=&D_{ij,\ell}\alpha +X_{ij,\ell}'\beta + R^Y_{ij,\ell} + \varepsilon_{ij,\ell},\:
\Ep[\varepsilon_{ij,\ell}|D_{ij,\ell},X_{ij,\ell}]=0,
\end{align*}
where $Y_{ij,\ell}$ denotes a measure of trust, $D_{ij,\ell}$ denotes an intensity measure of slave trade, $X_{ij,\ell}$ contains polynomial basis elements of age up to degree 10, ethnic fractionalization, gender, urban residence, occupation, religion, living conditions, the interactions of the polynomial basis of age with all the dummy variables, and country fixed effects, consisting of 597 dimensions of covariates in total.
Note that the total number of regressors ($p+1=598$) is much larger than the effective sample size ($\C = N \wedge M = 185$) of two-way clustering in this extended setting. 

Table \ref{tab:results_nunn} summarize the estimates of the effects of slave trade on mistrust as measured by the ``trust of neighbors,'' corresponding to Table 1 of \citet{NunnWantchekon11}.
The last two columns in the table show the original estimates obtained under the prototypical model by \citet[][Table 1]{NunnWantchekon11} and corresponding lasso estimates obtained under more flexible model specification by our method.
Across all the measures of slave exports, the original estimates and our lasso estimates are similar with similar levels of statistical significance.
These results demonstrate that, even for flexible model specifications entailing high-dimensional covariates,
the proposed method allows to produce qualitatively similar results without extensive loss of significance, and we can thus confirm further robustness of the main empirical findings by \citet{NunnWantchekon11}.

\subsection{Pre-Colonial Institutions and Regional Developments in Africa}\label{sec:pre colonial institutions and regional developments in africa}
\citet{Michalopoulos_Papaioannou2013} analyze the effects of pre-colonial institutions on contemporary regional developments in Africa, controlling for various population, locational and geographic covariates including population density, distance to capital, distance to sea coast, distance to border, water area, land area, elevation, land suitable for agriculture, ecological suitability, petrolium, and diamond mine as well as country fixed effects in their baseline model.
Estimates of these effects are obtained by running regressions with a sample that pools $n_{ij}$ populated pixels $\ell \in [n_{ij}]$ in ethnic group $i$ and country $j$ across the cells $(i,j) \in [N] \times [M]$ of $N (=93)$ ethnic groups and $M (=48)$ countries.
Standard errors are computed by the two-way cluster-robust method of \citet{CGM11} for the ethnic group and district as two ways of clustering.

With our proposed method that is applicable to both high-dimensional models and multi-way clustering, they could consider even more flexible model specifications, for example, allowing for interactions of all combinations of geographical covariates and locational covariates.
We present estimates with standard errors under such extended models with flexible specifications, demonstrate that qualitatively similar results continue to be obtained without substantial loss of statistical significance, and thus confirm further robustness of the main empirical findings by \citet{Michalopoulos_Papaioannou2013}.

Consider the model
\begin{align*}
Y_{ij,\ell}=&D_{ij,\ell}\alpha +X_{ij,\ell}'\beta + R^Y_{ij,\ell} + \varepsilon_{ij,\ell},\:
\Ep[\varepsilon_{ij,\ell}|D_{ij,\ell},X_{ij,\ell}]=0,
\end{align*}
where $Y_{ij,\ell}$ denotes a regional development measured by night light intensity, $D_{ij,\ell}$ denotes an intensity measure of pre-colonial ethinic institutions, $X_{ij,\ell}$ contains population density, interactions of all combinations of locational controls (distance to capital, distance to sea coast, and distance to border), interactions of all combinations of geographical controls (water area, land area, elevation, land suitable for agriculture, ecological suitability, petrolium, and diamond mine), and country fixed effects, consisting of 82 or 83 dimensions of covariates in total.
Note that the total number of regressors ($p+1=83$ or 84) is much larger than the effective sample size ($\C = N \wedge M = 48$) of two-way clustering in this extended setting. 

Table \ref{tab:results_michalopoulos} summarize the estimates of the effects of pre-colonial institutions on regional development as measured by the ``light density,'' corresponding to parts of Table 3 of \citet{Michalopoulos_Papaioannou2013}.
The last two columns in the table show the original estimates obtained under the prototypical model by \citet[][Table 3]{Michalopoulos_Papaioannou2013} and corresponding lasso estimates obtained under more flexible model specification by our method.
Across all the measures of pre-colonial institutions and all specifications, the original estimates and our lasso estimates are similar with similar levels of statistical significance.
These results demonstrate that, even for flexible model specifications entailing high-dimensional covariates,
the proposed method allows to produce qualitatively similar results without extensive loss of significance, and we can thus confirm further robustness of the main empirical findings by \citet{Michalopoulos_Papaioannou2013}.

\section{Conclusion}\label{sec:conclusion}
In this paper, we investigate high-dimensional regression models when data is sampled under multi-way clustering. 
We establish the convergence rates for the lasso and post-lasso estimators under multi-way clustering.
We then propose an inference method based on a post-double-selection procedure and show that it is asymptotically valid under multi-way clustering. 
Simulation studies demonstrate that the proposed procedure works well in finite sample under multi-way clustering in comparison with existing alternatives. 
We demonstrate that our method can enrich the flexibility of regression models and robustness of empirical results through a couple of empirical applications in growth and development economics.

Indeed, both multi-way clustering and high dimensionality are two important issues which concern applied research.
The existing literature provide solutions to each of multi-way clustering and high-dimensionality separately.
To our best knowledge, the literature does not seem to provide a solution to both of these issues simultaneously.
In this paper, we filled this void in the literature.

\newpage
\appendix
\section{Mathematical Proofs}

Throughout, the symbol $=$ stands for $\overset{a.s.}{=}$. 
We use the notations $Y=[Y_{11},...,Y_{NM}]'$, $X=[X_{11},...,X_{NM}]'$, $D=[D_{11},...,D_{NM}]'$, $\E=[\varepsilon_{11},...,\varepsilon_{NM}]'$, $V=[v_{11},...,v_{NM}]'$, $R^Y=[R^Y_{11},...,R^Y_{NM}]'$, $R^D=[R^D_{11},...,R^D_{NM}]'$, $g=X\beta + R^Y$, and $m=X\gamma + R^D$. 
For any $A\subset [p]$, let $X_A=\{X_j: j\in A\}$, where $X_j$ denotes the $j$-th the columns of $X$.
Also define the projection operator by
\begin{align*}
\mathcal P_A=X_A(X_A'X_A)^{-} X_A',
\end{align*}
and the orthogonal projection operator by $\mathcal M_A=I-\mathcal P_A$. 

\subsection{Proof of Lemma \ref{lemma:hajek}}\label{sec:lemma:hajek}

\begin{proof}
Our proof strategy closely follows that of Lemma D.2 in \cite{DDG18}, except that we care about allowing the data generating process to vary with $n$ to accommodate the increasing dimensionality $p$.
 
Under Assumption \ref{a:sampling} (1) and (2), Lemma C.1 (a version of Aldous-Hoover representation) of \cite{DDG18} implies that, for each $n$, there exists a measurable function $\tau_n$ such that
\begin{align}
\{Z_{ij}\}_{(i,j)\in \NN^2}=\{\tau_n(U_{i0},U_{0j},U_{ij} )\}_{(i,j)\in \NN^2} \label{eq:aldous-hoover_representation}
\end{align}
holds,
where $\left\{\{U_{i0}\}_{i\in\NN},\{U_{0j}\}_{j\in\NN},\{U_{ij}\}_{(i,j)\in\NN^2}\right\}$ are i.i.d. uniform$(0,1)$ random variables.

The H\'ajek projection $H_n f$ of $\GC f$ on the set $\G_n$ is characterized by
\begin{align*}
\Ep \Big[ (\GC f- H_n f)\cdot g(U_n)\Big]
=0\, \quad\text{ for any $g(U_n)\in \G_n$},
\end{align*}
where $U_n=(U_{i0},U_{0j})_{i\in[N],j\in[M]}$.
Thus, for any $U_c $ with $c=(c_1,c_2) \in \I_n=\{(i,0) ,(0,j):i\in [N],j\in[M]\}$, we have 
\begin{align*}
\Ep[\GC f|U_c]=\Ep[H_n f|U_c].
\end{align*}
Because the range of $H_n$ is a closed subspace, we have
\begin{align*}
H_n f= \sumi \Ep[H_n f|U_{i0}] + \sumj \Ep[H_n f|U_{0j}].
\end{align*}
It follows from the above two equations that
\begin{align*}
H_n f= \sumi \Ep[\GC f|U_{i0}] + \sumj \Ep[\GC f|U_{0j}].
\end{align*}

Now, fix $c\in\I_n$ and let $e(c)=(\mathbbm 1\{c_1>0\},\mathbbm 1\{c_2>0\})$. 
By the independence of $\left\{\{U_{i0}\}_{i\in\NN},\{U_{0j}\}_{j\in\NN},\{U_{ij}\}_{(i,j)\in\NN^2}\right\}$, $Z_{ij}$ and $U_c$ are independent whenever $c\ne (i,j)\odot e(c)$, where $\odot$ denots the Hadamard product. Thus, $ \Ep\Big[f(Z_{ij})- \Ep f(Z_{11}) \Big| U_c\Big]=\Ep\Big[f(Z_{ij})- \Ep f(Z_{11}) \Big]=0$.
Therefore,
\begin{align*}
\Ep[\GC f|U_c]=& 
\frac{\sqrt{\C}}{NM} \sumi\sumj  \Ep\Big[f(Z_{ij})- \Ep f(Z_{11}) \Big| U_c\Big]\\
=&
\frac{\sqrt{\C}}{NM} \sumi\sumj \1\{(i,j)\odot e(c) =c\} \Ep\Big[f(Z_{ij})- \Ep f(Z_{11}) \Big| U_c\Big].
\end{align*}
The representation (\ref{eq:aldous-hoover_representation}) implies, for all $(i,j)$ such that $(i,j) \odot e(c)=c$,
\begin{align*}
\Ep\Big[f(Z_{ij})- \Ep f(Z_{11}) \Big| U_c\Big]=&\Ep\Big[f(Z_{c\vee \textbf{1}})- \Ep f(Z_{11}) \Big| U_c\Big],
\end{align*}
i.e. the index outside the support of $c$ can be changed to $1$.
Now, suppose $c_k=0$, $k\in\{1,2\}$.
The representation (\ref{eq:aldous-hoover_representation}) again gives
\begin{align*}
\frac{\sqrt{\C}}{NM} \sumi\sumj \1\{(i,j)\odot e(c) =c\} \Ep\Big[f(Z_{c\vee \textbf{1}})- \Ep f(Z_{11}) \Big| U_c\Big]
=& 
\frac{\sqrt{\C}C_k}{NM}  \Ep\Big[f(Z_{c\vee \textbf{1}})- \Ep f(Z_{11}) \Big| U_c\Big],
\end{align*}
where $C_1=N$, $C_2=M$. 
Therefore,
\begin{align*}
H_n f=&\sum_{c \in \I_n} \frac{\sqrt{\C}C_k}{NM}  \Ep\Big[f(Z_{c\vee \textbf{1}})- \Ep f(Z_{11}) \Big| U_c\Big]\\
=&\sumi \frac{\sqrt{\C}}{N}  \Ep\Big[f(Z_{i1})- \Ep f(Z_{11}) \Big| U_{i0}\Big] + \sumj\frac{\sqrt{\C}}{M} \Ep\Big[f(Z_{1j})- \Ep f(Z_{11}) \Big| U_{0j}\Big],
\end{align*}
and each term in the two summands is independent from the others.
This establishes the first claim of the lemma.

We next show that the variance of $H_n f$ can be calculated as
\begin{align*}
V(H_n f)=&\mu_N V(\Ep[f(Z_{11})|U_{10}]) +  \mu_M  V(\Ep[f(Z_{11})|U_{01}])\\
=&\mu_N  Cov(f(Z_{11}),f(Z_{12})) + \mu_M Cov(f(Z_{11}),f(Z_{21})).
\end{align*}
To see this, note that
\begin{align*}
V(\Ep[f(Z_{11})|U_{10}])=&Cov(\Ep[f(Z_{11})|U_{10}],\Ep[f(Z_{12})|U_{10}])\\
=&Cov(f(Z_{11}),f(Z_{12}) ) -\Ep[ Cov(f(Z_{11}),f(Z_{12})|U_{10})]\\
=&Cov(f(Z_{11}),f(Z_{12}) ) -\Ep[ Cov\{f(\,\tau_n (U_{10},U_{01},U_{11}) \,),f(\,\tau_n (U_{10},U_{02},U_{12})\,) |U_{10}\}]\\
=&Cov(f(Z_{11}),f(Z_{12}) )-0,
\end{align*}
where 
the first equality follows from the representation (\ref{eq:aldous-hoover_representation}), 
the second from the law of total covariance, 
the third from the representation (\ref{eq:aldous-hoover_representation}), and  
the last from the fact that $\left\{\{U_{i0}\}_{i\in\NN},\{U_{0j}\}_{j\in\NN},\{U_{ij}\}_{(i,j)\in\NN^2}\right\}$ are independent. 
Analogous lines of calculations yield $V(\Ep[f(X_{11})|U_{01}])=Cov(f(X_{11}),f(X_{21}) )$.
Also, a direct calculation using Assumption \ref{a:sampling} (1) and (2) shows
\begin{align*}
V(\GC f)= &\frac{\C}{(NM)^2} \sumi \sum_{1\le j,j'\le M} Cov(f(Z_{ij}) ,f(Z_{ij'}) ) +
\frac{\C}{(NM)^2} \sum_{1\le i,i'\le N}\sumj Cov(f(Z_{ij}) ,f(Z_{i'j}) )\\
&-\frac{\C}{(NM)^2}\sumi\sumj V(f(Z_{ij}))\\
=&\mu_N  Cov(f(Z_{11}) ,f(Z_{12}) ) +
\mu_M Cov(f(Z_{11}) ,f(Z_{21}) ) +O\Big(\frac{1}{\C}\Big),
\end{align*}
since $\C/NM\le 1/\C$ and $\Ep f^2$ is bounded over $n$. 
This establishes the second claim of the lemma.
\end{proof}

\subsection{Proof of Theorem \ref{theorem:rates}}\label{sec:theorem:rates}

\begin{proof}
We will focusing on the result for $\widetilde\gamma$ since the results for $\widetilde \eta$ will follow analogously. 
The proof is divided into four steps. 
The conclusions from the first two steps give the rates for lasso. 
The third step provides bounds for the rates of the post-lasso in terms of convergence rates of lasso.
The fourth step provides the $\ell_2$-norm rate.

\noindent \textbf{Step 1.} 
The oracle inequality follows directly from Lemma 6 of \cite{BCCH12}, which is applicable under Assumption \ref{a:sparsity} (1)--(2). This implies that we have the following bounds for lasso estimator:
\begin{align*}
\|\hat\gamma  - \gamma\|_1 &\lesssim \frac{\sqrt{s} \lambda_2}{NM} + \frac{NM c_s^2}{\lambda_2 } \quad\text{and}\\
\|X_{ij}'(\hat\eta  - \eta)\|_n &\lesssim \frac{s \lambda_2}{NM} +c_s,
\end{align*}
conditionally on the regularized event $\lambda_2 /NM  \ge c \|(NM)^{-1}\sumi\sumj X_{ij}v_{ij}\|_\infty$ for some constant $c>1$.
Apply Assumption \ref{a:sparsity} (2) and use the choice of $\lambda_2$ to obtain the desired results in the first line.

\noindent \textbf{Step 2.}
We now claim that, if we set $\lambda_2 = O\Big(\sqrt{(NM)^2 \log a/\C}\Big)$, then the regularized event
\begin{align}
\max_{k \in [p]}\Big|\frac{1}{NM} \sumi \sumj X_{ij,k}v_{ij} -\Ep[X_{11,k}v_{11}]\Big| \lesssim \frac{1}{c}\sqrt{\frac{\log a}{\C}}=\frac{\lambda_2}{NM}. \label{eq:regularized_event}
\end{align}
realizes with probability at least $1-C(\log \C)^{-1}$.

First notice that the left-hand side can be bounded as
\begin{align*}
&\max_{k \in [p]}\Big|\frac{1}{NM} \sumi \sumj X_{ij,k}v_{ij} -\Ep[X_{11,k}v_{11}]\Big|\\
\le &
\max_{k \in [p]}\Big|\frac{1}{NM} \sumi \sumj X_{ij,k}v_{ij}-
\frac{M}{NM}\sumi 
\Ep[ X_{i1,k}v_{i1}|U_{i0}]
-
\frac{N}{NM}\sumj 
\Ep[ X_{1j,k}v_{1j}|U_{0j}] \Big|\\
&+\max_{k\in[p]}\Big|\frac{M}{NM}\sumi 
\Ep[ X_{i1,k}v_{i1}|U_{i0}]-\Ep[X_{11,k}v_{11}]\Big|
+\max_{k\in[p]}\Big|
\frac{N}{NM}\sumj 
\Ep[ X_{1j,k}v_{1j}|U_{0j}]
-\Ep[X_{11,k}v_{11}]\Big|\\
=&(1)+(2)+(3)
\end{align*}
where $E[X_{11,k}v_{11}]=0$ is used.
Part $(1)$ is $\Op(\frac{1}{\C})$ by Lemma \ref{lemma:hajek} under Assumptions \ref{a:sampling} and \ref{a:moments} (1). 
Using Lemma \ref{lemma:concentration_inequality_CCK} (see Appendix \ref{sec:auxiliary_lemmas} ahead), we can show $(2)=\Op(\sqrt{\frac{\log a}{N}})$. 
To see this, note that Assumption \ref{a:moments} (1) implies that $\sigma^2:=\max_{k\in[p]}\frac{1}{N}\sumi \Ep(\Ep[X_{i1}v_{i1}|U_{i0}])^2$ is uniformly bounded, and Assumption \ref{a:moments} (1)--(2) suggests
\begin{align*}
B^2=:&
\Ep[\max_{i\in [N]} \max_{k\in [p]} (\Ep[X_{i1.k}v_{i1}|U_{i0}])^2 ]\\
\le
&\Ep[\max_{i\in [N]} \|X_{i1}v_{i1}\|_\infty^2 ]\\
\le
&\Big(\Ep[\max_{i\in [N]} \|X_{i1}v_{i1}\|_\infty^{q} ]\Big)^{2/q}\\
\le
&N^{2/q}\Big(\Ep[ \|X_{11}\|_\infty^{q}|v_{11}|^{q} ]\Big)^{2/q}\\
\le
&N^{2/q}\Big(\sqrt{\Ep[ \|X_{11}\|_\infty^{2q}]} \sqrt{\Ep|v_{11}|^{2q} }]\Big)^{2/q}
=N^{2/q} B^2_n O(1),
\end{align*}
where the first inequality follows from the property of projection, the second inequality follows from Jensen's inequality, and the fourth inequality follows from Cauchy-Schwartz inequality.
Thus, Lemma \ref{lemma:concentration_inequality_CCK} (see Appendix \ref{sec:auxiliary_lemmas} ahead) implies
\begin{align*}
|(2)|\lesssim_\Pr \sqrt{\frac{\log a}{N}} + \frac{B_n \log a}{N^{1-1/q}}\lesssim\sqrt{\frac{\log a}{N}}.
\end{align*}
Finally, $(3)=\Op(\sqrt{\frac{\log a}{ M}})$ follows analogously.
Therefore, $(2)+(3)=\Op(\sqrt{\frac{\log a}{\C}})$. 

\noindent \textbf{Step 3.}
We now derive bounds for performance of post-lasso:
\begin{align}\label{eq:convergence_rate_step3_1}
\|m_{ij}-X_{ij} \widetilde \gamma\|_n \lesssim&_\Pr \sqrt{\frac{s\log a}{\C}} + \frac{ \|(I-\PP_{\hat I_2})m \| }{ \sqrt{NM} }
\end{align}
where $m_{ij}=X'_{ij}\gamma + R^D_{ij}$. 
This part of proof closely follows the proof of Lemma 7 in \cite{BCCH12} with some minor modifications. 
First note that
$
m-X_{\hat I_2} \widetilde \gamma=(I-\PP_{\hat I_2})m -\PP_{\hat I_2}V.
$
This implies 
\begin{align*}
\|m-X_{\hat I_2} \widetilde\gamma\| \le& \|(I-\PP_{\hat I_2})m \|+\| \PP_{\hat I_2}V\|.
\end{align*}
By the definition of $\PP_{\hat I_2}$ and the operator norm,
$$
\| \PP_{\hat I_2}V\|\le \|X_{\hat I_2}/\sqrt{NM} (X_{\hat I_2}'X_{\hat I_2}/NM)^{-1}\|\,\|X_{\hat I_2}'V/\sqrt{MM}\|
$$ 
and 
\begin{align*}
 \|X_{\hat I_2}/\sqrt{NM} (X_{\hat I_2}'X_{\hat I_2}/NM)^{-1}\|\le \sqrt{1/\semin{s + \widetilde m_2}},
\end{align*}
where $\widetilde m_2=|\hat I_2 \setminus T_2|$, $T_2= \supp(\gamma)$.
Thus under Assumption \ref{a:sparse_eigenvalues}, we obtain
\begin{align*}
\| \PP_{\hat I_2}V\|\le& \sqrt{1/\semin{s + \widetilde m_2}} \,\|X_{\hat I_2}'V/\sqrt{MM}\| \le \sqrt{\frac{s+\widetilde m_2}{\semin{s + \widetilde m_2}}} \, \|XV/\sqrt{NM}\|_\infty\lesssim_\Pr  \sqrt{\frac{s\log a}{\C}},
\end{align*}
where the last inequality follows from equation (\ref{eq:regularized_event}) and Lemma \ref{lemma:empirical_pre-sparsity} (see Appendix \ref{sec:auxiliary_lemmas} ahead).
This shows (\ref{eq:convergence_rate_step3_1}).

By (\ref{eq:regularized_event}), the same argument as that of Lemma 7 in \cite{BCCH12} establishes
\begin{align*}
\frac{ \|(I-\PP_{\hat I_2})m \| }{ \sqrt{NM} } \lesssim& \frac{\sqrt{s} \lambda_2}{NM} + c_s.
\end{align*}
Therefore, (\ref{eq:convergence_rate_step3_1}) can be rewritten as
\begin{align*}
\|m_{ij}-X_{ij} \widetilde \gamma\|_n \lesssim&_\Pr \sqrt{\frac{s\log a}{\C}} +  \frac{\sqrt{s} \lambda_2}{NM} + c_s.
\end{align*}
Next, applying Lemma \ref{lemma:empirical_pre-sparsity} (see Appendix \ref{sec:auxiliary_lemmas} ahead), we have
\begin{align*}
\|\widetilde \gamma - \gamma\|_1
\le 
\sqrt{\|\widetilde \gamma - \gamma\|_0}\, \|\widetilde \gamma - \gamma\|
\le  \sqrt{s+\widetilde m_2}\, \frac{\|X_{ij}(\widetilde \gamma - \gamma)\|_n}{\sqrt{\semin{s+\widetilde m_2}}}
\end{align*}
under Assumption \ref{a:sparsity} (1) and \ref{a:sparse_eigenvalues}.

Combining the above bounds, the choice of $\lambda_2$, and Assumption \ref{a:sparsity} (2), we obtain
\begin{align*}
\|\widetilde \gamma - \gamma\|_1\lesssim&_\Pr  \sqrt{\frac{s^2\log a}{\C}}
\qquad\text{and}
\\
\|X_{ij} (\widetilde \gamma - \gamma)\|_n\lesssim&_\Pr  \sqrt{\frac{s\log a}{\C}}.
\end{align*}

\noindent \textbf{Step 4.}
The $\ell_2$-norm rates are implied by the prediction norm rates, Assumption \ref{a:sparse_eigenvalues}, and Lemma \ref{lemma:empirical_pre-sparsity} (see Appendix \ref{sec:auxiliary_lemmas} ahead).
\end{proof}

\subsection{Proof of Theorem \ref{theorem:asymptotic_normality}}\label{sec:theorem:asymptotic_normality}

\begin{proof}
Our proof follows parallel steps to Steps 1-6 in the proof of Theorem 1 in \citet{BCH14}. 
However, due to the two-way cluster sampling, most of the probabilistic bounds are established differently.

We use the short-hand notation
\begin{align*}
\widetilde b_Z(A):=\underset{b\in \Real^p: b_j=0 \forall j\in A^c}{\argmin} \|Z-X'b\|^2
\end{align*}
for any vector $Z\in \Real^n$.

\noindent\textbf{Step 1}
Write $\widetilde \alpha=[D'\MM_{\hat I}D/NM]^{-1} D'\MM_{\hat I}Y/NM$ and thus we can write
\begin{align*}
\sqrt{\C} (\widetilde \alpha - \alpha)=[D'\MM_{\hat I}D/NM]^{-1} \cdot\sqrt{\C}D'\MM_{\hat I}(g+\E)/NM=:(II)^{-1} \cdot (I).
\end{align*}
By Steps 2 and 3 to be presented below, we obtain
\begin{align*}
(II)= V'V/NM +\op(1) \text{ and } (I)= \sqrt{\C}V'\E/NM +\op(1).
\end{align*}
Also note that $V'V/NM =\Ep[v_{11}^2]+\op(1)$ by Lemma \ref{lemma:hajek} and Assumption \ref{a:sampling}--\ref{a:moments}, which can be shown following the same arguments as those in Step 3 of the proof for Theorem \ref{theorem:variance_est}.
Under Assumption \ref{a:moments} (1), $\Ep[v_{11}^2]$ is bounded and bounded away from zero uniformly in $n$.
Therefore $(II)^{-1}=\Ep[v_{11}^2]^{-1}+\op(1)$. 

Under Assumption \ref{a:moments} (3), $\sigma^2$ is bounded and bounded away from zero. 
Setting $W_{ij}:=\sigma^{-1}  v_{ij}\varepsilon_{ij}$ and $Z_{ij}\overset{f}{\mapsto} W_{ij}$, we have $\Ep f(Z_{11})=0$ and
\begin{align*}
\GC f=\frac{\sqrt{\C}}{NM}\sumi\sumj W_{ij}=\sigma^{-1} \sqrt{\C} (\widetilde \alpha -\alpha)+\op(1).
\end{align*}
Apply Lemma \ref{lemma:hajek} under Assumption \ref{a:sampling} and \ref{a:moments} (1) to obtain the H\'ajek projection 
\begin{align*}
H_n f
=&\sumi \frac{\sqrt{\C}}{N}  \Ep [f(Z_{i1}) | U_{i0}] + \sumj\frac{\sqrt{\C}}{M} \Ep [f(Z_{1j}) | U_{0j}]
\end{align*}
of $\GC f$,
where terms in each summand are independent and two summands are independent of each other. 
We now check Lyapunov's conditions.
First, note that Assumption \ref{a:moments} (1) guarantees that the third moments of both summands are bounded uniformly in $n$. 
Second, the second part of Lemma \ref{lemma:hajek} implies that
\begin{align*}
\lim_{n \rightarrow \infty} V(H_n f)=
\bar\mu_N V(\Ep [f(Z_{11}) | U_{10}])  + \bar\mu_M  V( \Ep [f(Z_{11}) | U_{01}]) 
\\
= \bar \mu_N \Ep[ f(Z_{11})f(Z_{12})] + \bar \mu_M \Ep[ f(Z_{11})f(Z_{21})]
=\Gamma \in (c,\infty)
\end{align*}
a.s. for $c > 0$,
where the last inequalities follow from Assumption \ref{a:moments} (3).
Therefore, we apply Lyapunov's CLT to obtain
\begin{align*}
H_n f \leadsto N\Big(0, \bar\mu_N V(\Ep [f(Z_{11}) | U_{10}])  + \bar\mu_M  V( \Ep [f(Z_{11}) | U_{01}])  \Big).
\end{align*}

The first equality in the variance equation of Lemma \ref{lemma:hajek} yields
\begin{align*}
V(\GC f)= \bar \mu_N \Ep[ f(Z_{11})f(Z_{12})] + \bar \mu_M \Ep[ f(Z_{11})f(Z_{21})] + \op(1),
\end{align*}
where the right-hand side is asymptotically positive and bounded away from zero. Therefore,
\begin{align*}
\sqrt{\C}(\widetilde \alpha - \alpha) = \GC f \leadsto N(0,\sigma^2).
\end{align*}
 
\noindent\textbf{Step 2}
Use $D=m+V$ to decompose  
\begin{align*}
(I)=&
\sqrt{\C}V'\E/NM 
+
 \sqrt{\C}m'\MM_{\hat I}g/NM 
+
\sqrt{\C}m'\MM_{\hat I}\E/NM
+
\sqrt{\C}V'\MM_{\hat I}g/NM
-
\sqrt{\C}V'\PP_{\hat I}\E/NM\\
=&
\sqrt{\C}V'\E/NM 
+
(1a)
+
(1b)
+
(1c)
-
(1d).
\end{align*}
By Steps 5 and 6 to be presented below, we have
\begin{align*}
|(1a)|\lesssim  \sqrt{\C}\|\MM_{\hat I}m/\sqrt{NM}\|\cdot\|\MM_{\hat I}g/\sqrt{NM}\|\lesssim_\Pr \sqrt{\frac{s^2 (\log a)^2}{\C}}.
\end{align*}
Using the decompositions $m=X\gamma + R^D$, $m'\PP_{\hat I}=\widetilde b'_m(\hat I) X'$ and $m'\MM_{\hat I} \E= (R^D)'\E - (\widetilde b_m(\hat I) - \gamma)'X' \E$, one has
\begin{align*}
|(1b)|\le \sqrt{\C}|R^{D\prime}\E/NM| + \sqrt{\C}|(\widetilde b_m(\hat I)- \gamma )X'\E/NM|\lesssim_\Pr  \sqrt{\frac{s^2 (\log a)^2}{\C}},
\end{align*}
because, under Assumptions \ref{a:moments} (1) and \ref{a:sparsity} (2),
\begin{align*}
\sqrt{\C}|R^{D\prime}\E/NM|\le \sqrt{\C}\sqrt{R^{D\prime} R^D/NM} \cdot \Op\Big(\sqrt{\frac{1}{NM}\Ep\|\E\|^2}\Big) \lesssim_\Pr \sqrt{\frac{s }{\C}}
\end{align*}
and
\begin{align*}
\sqrt{\C}|(\widetilde b_m(\hat I)- \gamma )X'\E/NM| \le \sqrt{\C}\|\widetilde b_m (\hat I) - \gamma\|_1 \|X'\E/NM\|_\infty  
\\
\lesssim_\Pr \sqrt{\C} \sqrt{\frac{s^2 \log a}{\C}} \cdot \sqrt{\frac{\log a}{\C}}=\sqrt{\frac{s^2 (\log a)^2}{\C}},
\end{align*}
where $\|\widetilde b_m (\hat I) - \gamma\|_1\lesssim_\Pr \sqrt{\frac{s^2 \log a}{\C}}$ follows from Step 5 and $\|X'\E/NM\|_\infty \lesssim_\Pr \sqrt{\frac{ \log a}{\C}}$ follows from Step 4.
Third, using the same argument as above for (1b) following Steps 4 and 6 and $g=X\beta+R^Y$, we have
\begin{align*}
|(1c)|\le \sqrt{C}|R^{Y\prime}V/\sqrt{NM}| + \sqrt{\C} |(\widetilde b_g (\hat I)-\beta)'X'V/\sqrt{NM}|
\lesssim_\Pr \sqrt{\frac{s}{\C}} + \sqrt{\frac{s^2 (\log a)^2}{\C}}.
\end{align*}
Finally, 
\begin{align*}
|(1d)|\le \sqrt{\C}|\widetilde b_V (\hat I)'X'\E /NM |\le \sqrt{\C}\|\widetilde b_V (\hat I)\|_1\|X'\E /NM \|_\infty \lesssim_\Pr \sqrt{\frac{s^2 (\log a)^2 }{\C}}
\end{align*}
following equation (\ref{eq:regularized_event}) in the proof of Theorem \ref{theorem:rates}, and 
\begin{align*}
\|\widetilde b_V (\hat I)\|_1 \le \sqrt{\hat s} \|\widetilde b_V (\hat I)\| \le \sqrt{\hat s}\|(X'_{\hat I} X_{\hat I}/NM)^{-1} X'_{\hat I} V/NM \|\\
\lesssim_\Pr \frac{\sqrt{\hat s}}{\semin{C\hat s}}\sqrt{\hat s}\| X'_{\hat I} V/NM \|_\infty\lesssim_\Pr  \sqrt{\frac{s^2 \log a}{\C}}
\end{align*}
following Step 4, Lemma \ref{lemma:empirical_pre-sparsity}, and Assumption \ref{a:sparse_eigenvalues}.

\noindent\textbf{Step 3}
We can write
\begin{align*}
(II)=&(m+V)'\MM_{\hat I} (m+V)/NM\\
=&V'V/NM + m'\MM_{\hat I}m/NM +2 m'\MM_{\hat I}V/NM -V'\PP_{\hat I}V/NM\\
=&V'V/NM +(2a) +(2b) - (2c).
\end{align*}
We have 
$|(2a)|\lesssim_\Pr \sqrt{\frac{s^2(\log a)^2}{\C}}$ by Step 5, 
$|(2b)|\lesssim_\Pr \sqrt{\frac{s^2(\log a)^2}{\C}}$ by a similar argument to bounding $|(1b)|$, and 
$|(2c)|\lesssim_\Pr \sqrt{\frac{s^2(\log a)^2}{\C}}$ by a similar argument to bounding $|(1d)|$.

\noindent\textbf{Step 4}
In this step, we show that the following regularized events hold with probability $1-o(1)$:
\begin{align*}
(a)\: \sqrt{\C}\|X'\E/NM\|_\infty \lesssim_\Pr \sqrt{\log a} \text{ and } (b) \: \sqrt{\C}\|X'V/NM\|_\infty \lesssim_\Pr \sqrt{\log a}.
\end{align*}
This claim follows from similar lines of argument to those showing equation (\ref{eq:regularized_event}) in the proof of Theorem \ref{theorem:rates} under Assumptions \ref{a:sampling} and \ref{a:moments} (1)--(2).

\noindent\textbf{Step 5}
In this step, we show 
\begin{align*}
(a)\: \|\MM_{\hat I}m/\sqrt{NM}\| \lesssim_\Pr \sqrt{\frac{s \log a}{\C}} \text{ and } (b) \: \|\widetilde b_m(\hat I)-\gamma\| \lesssim_\Pr \sqrt{\frac{s \log a}{\C}}.
\end{align*}
First, by applying Theorem \ref{theorem:rates} under Assumptions \ref{a:sampling}, \ref{a:moments} (1)--(2), \ref{a:sparsity}(1)--(2), and \ref{a:sparse_eigenvalues}, and by following the same argument as the one in Step 5 of \cite{BCH14}, we have
\begin{align*}
\|\MM_{\hat I} m/\sqrt{NM}\|\le\|\MM_{\hat I_2} m/\sqrt{NM}\|\le\|(X\widetilde b_D(\hat I_2) -m)/\sqrt{NM}\|
\lesssim_\Pr  \sqrt{\frac{s \log a}{\C}},
\end{align*}
where the first inequality follows from $\hat I_2 \subset \hat I$ and the second follows from the fact that $\widetilde b_m(\hat I_2)$ minimizes $\|m-X'_{\hat I_2}b\|$.
Second, the reverse triangle inequality yields
\begin{align*}
\Big|\|X(\widetilde b_m (\hat I) -\gamma )/\sqrt{NM}\| - \|R_m/ \sqrt{NM}\|\Big|\lesssim_\Pr\|\MM_{\hat I} m/\sqrt{NM}\|,
\end{align*}
and, by Assumption \ref{a:sparsity} (2), $\|R_m/ \sqrt{NM}\|\lesssim_\Pr \sqrt{s/\C}$. 
Thus, by using Lemma \ref{lemma:empirical_pre-sparsity} with Assumptions \ref{a:sparsity} (1) and \ref{a:sparse_eigenvalues}, we obtain
\begin{align*}
\|\widetilde b_m (\hat I) - \gamma\|\lesssim&_\Pr \sqrt{\semin{\hat s + s}} \|\widetilde b_m (\hat I) - \gamma\|\\
\le& \|X(\widetilde b_m (\hat I) - \gamma)/\sqrt{NM}\|
\lesssim_\Pr \sqrt{\frac{s \log a}{\C}}.
\end{align*}

\noindent\textbf{Step 6}
Finally, we can show 
\begin{align*}
(a)\: \sqrt{\C}\|\MM_{\hat I}g/NM\| \lesssim_\Pr \sqrt{\frac{s \log a}{\C}} \text{ and } (b) \: \|\widetilde b _g(\hat I)-\beta\| \lesssim_\Pr \sqrt{\frac{s \log a}{\C}}.
\end{align*}
following similar lines of argument to those of Step 5 under Assumptions \ref{a:sampling}, \ref{a:moments}, \ref{a:sparsity}, and \ref{a:sparse_eigenvalues}.
\end{proof}

\subsection{Proof of Theorem \ref{theorem:variance_est}}\label{sec:theorem:variance_est}

\begin{proof}
First, note that we have the following decomposition
\begin{align*}
&|\hat Q^{-1} \hat \Gamma \hat Q^{-1}- Q^{-1}  \Gamma  Q^{-1}|\\
\lesssim & |\hat Q^{-1}-Q^{-1}||\hat Q^{-1}+Q^{-1}||\hat \Gamma| + |\hat \Gamma - \Gamma||Q^{-1}|^2,
\end{align*}
where $|Q^{-1}|$ is bounded away from zero uniformly by Assumption \ref{a:moments}(1). The rest of this proof is divide into 5 steps.
In Steps 1 and 2, we obtain a bound for $|\hat \Gamma-\Gamma|$. 
In Steps 3 and 4, we obtain a bound for $|\hat Q^{-1} - Q^{-1}|$. 
Finally, Step 5 shows a bound for $|\hat \Gamma|$ and $|\hat Q^{-1}+ Q^{-1}|$.

\noindent\textbf{Step 1.} 
We derive a bound for $|\widetilde \Gamma - \Gamma|$, where
\begin{align*}
\widetilde \Gamma=  \frac{\mu_N}{NM^2}\sumi\sum_{1\le j,j'\le M}  v_{ij} \varepsilon_{ij}  v_{ij'}   \varepsilon_{ij'}  +  \frac{\mu_M}{N^2 M}\sum_{1\le i,i'\le N}\sumj  v_{ij} \varepsilon_{ij} v_{i'j}  \varepsilon_{i'j}.
\end{align*}
We first claim that
\begin{align}
 \frac{\mu_N}{NM^2}\sumi\sum_{1\le j,j'\le M}  v_{ij} \varepsilon_{ij}  v_{ij'}   \varepsilon_{ij'} =\bar\mu_N\Ep[v_{11}\varepsilon_{11}v_{12}\varepsilon_{12}]+\op(1).\label{eq:variance_consistency_exchangeable}
\end{align}

Note that, for each $n$, for any $i,\iota\in [N]$ and $j,k,l,m\in [M]$, we have
\begin{align*}
Cov\Big(
v_{ij} \varepsilon_{ij}  
v_{ik}   \varepsilon_{ik},
v_{\iota l} \varepsilon_{\iota l}  
v_{\iota m}   \varepsilon_{\iota m}
\Big)
\le& \max_{i\in [N],j,j'\in [M]} V(v_{ij} \varepsilon_{ij}  
v_{ij'}   \varepsilon_{ij'})\\
=&\max_{i\in [N],j,j'\in [M]}\Big\{\Ep[(v_{ij} \varepsilon_{ij}  
v_{ij'}   \varepsilon_{ij'})^2]-(\Ep[v_{ij} \varepsilon_{ij}  
v_{ij'}   \varepsilon_{ij'}])^2\Big\}.
\end{align*}
Using Cauchy-Schwartz's inequality with Assumptions \ref{a:sampling} (1) and \ref{a:moments} (1), the first term in the variance can be bounded as
\begin{align*}
\Ep[(v_{ij} \varepsilon_{ij}  
v_{ij'}   \varepsilon_{ij'})^2]
\le &
\sqrt{\Ep[v_{ij}^4 \varepsilon_{ij}^4 ] \Ep [v_{ij'}^4 \varepsilon_{ij'}^4]}\\
\le & 
\sqrt{\Ep v_{11}^8\Ep \varepsilon_{11}^8 }=O(1)
\end{align*}
uniformly over $n$.
Under Assumptions \ref{a:sampling} (1) and \ref{a:moments} (1), similar calculations can be carried out to the square-root of the second term to obtain
\begin{align*}
\Ep[v_{ij} \varepsilon_{ij}  
v_{ij'}   \varepsilon_{ij'}]\le & \sqrt{\Ep[v_{ij}^2 \varepsilon_{ij}^2  ]\Ep[v_{ij'}^2 \varepsilon_{ij'}^2  ]}\le \sqrt{\Ep[v_{11}^4 ]\Ep[\varepsilon_{11}^4    ]}=O(1)
\end{align*}
uniformly over $n$. 
This shows that, for any $n$, for any $i,\iota\in [N]$ and $j,k,l,m\in [M]$, it holds that, for a $K>0$ independent of $n$,
\begin{align}
\Big|Cov\Big(
v_{ij} \varepsilon_{ij}  
v_{ik}   \varepsilon_{ik},
v_{\iota l} \varepsilon_{\iota l}  
v_{\iota m}   \varepsilon_{\iota m}
\Big)\Big|\le K.\label{eq:covariance}
\end{align}

With this bound of the covariance, we can bound the variance as
\begin{align*}
&V\Big(\frac{1}{NM^2}\sumi\sum_{1\le j,j'\le M}  v_{ij} \varepsilon_{ij}  v_{ij'}   \varepsilon_{ij'} \Big)
\\
=&Cov\Big(\frac{1}{NM^2}\sumi\sum_{1\le j,j'\le M}  v_{ij} \varepsilon_{ij}  v_{ij'}   \varepsilon_{ij'},\frac{1}{NM^2}\sumi\sum_{1\le j,j'\le M}  v_{ij} \varepsilon_{ij}  v_{ij'}   \varepsilon_{ij'} \Big)\\
=&\frac{1}{N^2 M^4}\sumi \sum_{1\le j,k,l,m\le M} Cov\Big(
v_{ij} \varepsilon_{ij}  
v_{ik}   \varepsilon_{ik},
v_{il} \varepsilon_{il}  
v_{im}   \varepsilon_{im}
\Big)\\
&+
\frac{2}{N^2 M^4}\sumj \sum_{1\le i, i'\le N} \sum_{1\le k,l \le M} Cov\Big(
v_{ij} \varepsilon_{ij}  
v_{ik}   \varepsilon_{ik},
v_{i'j} \varepsilon_{i'j}  
v_{i'l}   \varepsilon_{i'l}
\Big)+o\Big(\frac{1}{\C}\Big)\\
=&O\Big(\frac{1}{\C}\Big)=o(1)
\end{align*}
uniformly over $n$, where the second equality follows from Assumption \ref{a:sampling} (2) and counting the number of terms in each summand, and the third equality is due to (\ref{eq:covariance}). 
Applying Chebyshev's inequality, it follows that 
\begin{align}
&\Pr\Big(\Big|\frac{1}{NM^2}\sumi\sum_{1\le j,j'\le M} ( v_{ij} \varepsilon_{ij}  v_{ij'}   \varepsilon_{ij'}- \Ep[ v_{ij} \varepsilon_{ij}  v_{ij'}   \varepsilon_{ij'}])\Big|>\epsilon\Big)
\nonumber\\
\le&
 \frac{\sup_n V\Big(\frac{1}{NM^2}\sumi\sum_{1\le j,j'\le M}  v_{ij} \varepsilon_{ij}  v_{ij'}   \varepsilon_{ij'}\Big)}{\epsilon^2}
 = \frac{1}{\epsilon^{2}}\cdot o(1).\label{eq:variance_consistency}
\end{align}
for any $\epsilon>0$.

Also, under Assumption \ref{a:moments} (1), the first result in Lemma D.10 of \cite{DDG18} ensures
\begin{align*}
\Ep\Big[\Big|\frac{\mu_N}{NM^2}\sumi\sum_{1\le j,j'\le M}  v_{ij} \varepsilon_{ij}  v_{ij'}   \varepsilon_{ij'} 
- 
\frac{\mu_N}{N M (M-1)}\sumi\sumj \sum_{j'\ne j}  v_{ij} \varepsilon_{ij}  v_{ij'}   \varepsilon_{ij'}\Big|\Big]=o(1)
\end{align*}
uniformly over $n$.
Furthermore, under Assumptions \ref{a:sampling} (1) and \ref{a:moments} (1), we have
\begin{align*}
\Ep\Big[\frac{1}{N M (M-1)}\sumi\sumj \sum_{j'\ne j}  v_{ij} \varepsilon_{ij}  v_{ij'}   \varepsilon_{ij'} \Big]=\Ep[v_{11}\varepsilon_{11} v_{12}\varepsilon_{12}].
\end{align*} 
Combining these with (\ref{eq:variance_consistency}), we obtain (\ref{eq:variance_consistency_exchangeable}).
A symmetric argument also shows 
$$
\frac{\mu_M}{N^2 M}\sum_{1\le i,i'\le N}\sumj  v_{ij} \varepsilon_{ij} v_{i'j}  \varepsilon_{i'j}
 =\bar \mu_M \Ep[v_{11}\varepsilon_{11} v_{21}\varepsilon_{21}]+\op(1).
$$
Therefore, we obtain $|\widetilde \Gamma - \Gamma|=\op(1)$.

\noindent\textbf{Step 2.} In this step we bound $|\hat \Gamma - \widetilde \Gamma|$, where $\widetilde \Gamma$ is defined in Step 1. 
Consider the decomposition
\begin{align*}
& \underbrace{\frac{\mu_N}{NM^2}\sumi\sum_{1\le j,j'\le M} \Big( 
\hat v_{ij} \hat\varepsilon_{ij} \hat v_{ij'}   \hat\varepsilon_{ij'}-
v_{ij} \varepsilon_{ij}  v_{ij'}   \varepsilon_{ij'}\Big) }_{(1)}
 + 
\\
&
\underbrace{ \frac{\mu_N}{N^2 M}\sum_{1\le i,i'\le N}\sumj 
\Big(
 \hat v_{ij}\hat \varepsilon_{ij}\hat v_{i'j}\hat  \varepsilon_{i'j}
 -
 v_{ij} \varepsilon_{ij} v_{i'j}  \varepsilon_{i'j}\Big)}_{(2)}.
\end{align*}
Recall
$
 \varepsilon_{ij}= Y_{ij} - Z'_{ij}\eta - R^Y_{ij},\:
 \hat \varepsilon_{ij}=Y_{ij}- Z'_{ij}\widetilde\eta,
$
$
v_{ij}= D_{ij} - X'_{ij}\gamma - R^D_{ij},\:
\hat v_{ij}= D_{ij} - X'_{ij}\widetilde\gamma,
$
and thus,
$
\hat\varepsilon_{ij}-\varepsilon_{ij}= Z'_{ij}(\widetilde\eta-\eta)
-R^Y_{ij}$ and $
\hat v_{ij}-v_{ij}= X'_{ij}(\widetilde\gamma-\gamma) - R^D_{ij}.
$
We can further decompose (1) as
\begin{align*}
(1)=& \frac{\mu_N}{NM^2}\sumi\sum_{1\le j,j'\le M} \Big( 
(\hat v_{ij}- v_{ij}) \hat\varepsilon_{ij}  \hat v_{ij'}   \hat\varepsilon_{ij'}\Big) 
+
 \frac{\mu_N}{NM^2}\sumi\sum_{1\le j,j'\le M} \Big( 
 v_{ij} (\hat\varepsilon_{ij}-\varepsilon_{ij}) \hat v_{ij'}   \hat\varepsilon_{ij'}\Big) \\
&+
 \frac{\mu_N}{NM^2}\sumi\sum_{1\le j,j'\le M} \Big( 
 v_{ij} \varepsilon_{ij} (\hat v_{ij'}-v_{ij'})   \hat\varepsilon_{ij'}\Big) 
+
 \frac{\mu_N}{NM^2}\sumi\sum_{1\le j,j'\le M} \Big( 
v_{ij} \varepsilon_{ij}  v_{ij'}   (\hat \varepsilon_{ij'}-\varepsilon_{ij'})\Big) \\
=&(1a)+(1b)+(1c)+(1d).
\end{align*}
Under Assumption \ref{a:moments} (1), we first bound
\begin{align*}
(1a)\lesssim_\Pr 
&\Big|\frac{\mu_N }{NM^2}\sumi\sum_{1\le j,j'\le M} \Big( 
(\hat v_{ij}-v_{ij}) \hat\varepsilon_{ij} \hat v_{ij'}  \hat \varepsilon_{ij'}\Big)\Big|\\
\le&
\Big| \frac{\mu_N}{NM^2}\sumi\sum_{1\le j,j'\le M} 
X_{ij}'(\hat \gamma-\gamma) \hat \varepsilon_{ij} \hat v_{ij'}  \hat \varepsilon_{ij'}\Big|
+
\Big| \frac{\mu_N}{NM^2}\sumi\sum_{1\le j,j'\le M} R^D_{ij} \hat\varepsilon_{ij} \hat v_{ij'} \hat  \varepsilon_{ij'}\Big|\\
=&(1aa)+(1ab).
\end{align*}
We obtain
\begin{align*}
(1aa)=&
\Big| \frac{\mu_N}{NM^2}\sumi\sum_{1\le j,j'\le M} 
X_{ij}'(\hat \gamma -\gamma)\hat\varepsilon_{ij} \hat v_{ij'}  \hat \varepsilon_{ij'}\Big|\\
\le &
\Big| \frac{\mu_N}{NM^2}\sumi\sum_{1\le j,j'\le M} 
X_{ij}'(\hat \gamma -\gamma)\cdot\Big(Z_{ij}'(\hat \eta - \eta) +\varepsilon_{ij} - R^Y_{ij}\Big)\\
&\qquad \cdot \Big(X_{ij'}'(\hat \gamma - \gamma) +v_{ij'} - R^D_{ij'}\Big)\cdot\Big(Z_{ij'}'(\hat \eta - \eta) +\varepsilon_{ij'} - R^Y_{ij'}\Big)\Big|
= \op(1),
\end{align*}
where 
the last equality follows from triangle inequality, Cauchy-Schwartz's inequality, Theorem \ref{theorem:rates}, Assumptions \ref{a:moments} (1)--(2) and \ref{a:sparsity} (2), and the rate conditions in the statement of the theorem.
To see this, note that, under Assumption \ref{a:moments} (2), Theorem \ref{theorem:rates}, and the rate condition in the theorem, we have
\begin{align*}
&\sqrt{
\frac{\C}{N^2 M^2}  
\sumi \sum_{1\le j,j' \le M}
\Big(X_{ij}' (\hat \gamma - \gamma) Z_{ij}'(\hat \eta - \eta)\Big)^2
}\\
\le& 
\sqrt{
\frac{\C}{N M} \max_{i\in [N],j\in [M]}\| Z_{ij}\|^2_\infty \|\hat \eta - \eta\|^2_1
M \|X_{ij}'(\hat \gamma -\gamma)\|_n^2
}\\
\lesssim&_\Pr 
\sqrt{
\frac{(NM)^{1/q} B_n^2 s^3 (\log a)^2}{\C N}
}=o(1).
\end{align*}
Furthermore, by Theorem \ref{theorem:rates} and Assumptions \ref{a:moments} (1) and \ref{a:sparsity} (2), we have
\begin{align*}
&\sqrt{
\frac{\C}{N^2 M^2}  
\sumi \sum_{1\le j,j' \le M}
\Big(X_{ij}' (\hat \gamma - \gamma) \varepsilon_{ij}\Big)^2
}\\
\le& 
\sqrt{
\frac{\C}{N M} \max_{i\in [N],j\in [M]}|\varepsilon_{ij}|^2
M \|X_{ij}'(\hat \gamma -\gamma)\|_n^2
}\\
\lesssim&_\Pr 
\sqrt{
\frac{(NM)^{1/q}  s \log a}{ N}
}=o(1).
\end{align*}
The rest of the terms can be shown to be of smaller orders using similar arguments.
Finally, the rate condition from the statement of the theorem gives
\begin{align*}
&\sqrt{
\frac{\C}{N^2 M^2}  
\sumi \sum_{1\le j,j' \le M}
\Big(R^D_{ij} R^Y_{ij}\Big)^2
}
\le 
\sqrt{
\|R^D_{ij}R^Y_{ij}\|^2_n
}
=O(1).
\end{align*}
Similarly, using Cauchy-Schwartz's inequality, Theorem \ref{theorem:rates}, Assumptions \ref{a:moments} (1)--(2) and \ref{a:sparsity} (2), and the addition rate conditions in the statement of the theorem,  we obtain
\begin{align*}
(1ab)=&
\Big| \frac{\mu_N}{NM^2}\sumi\sum_{1\le j,j'\le M} 
R^D_{ij}\hat\varepsilon_{ij} \hat v_{ij'}  \hat \varepsilon_{ij'}\Big|
=\op(1).
\end{align*}
These results yield $(1a)=\op(1)$.
Following analogous but simpler arguments,
we can show that $(1b)$, $(1c)$ and $(1d)$ are $\op(1)$. This shows $(1)=\op(1)$. 
Similar lines of argument under the same set of assumptions show $(2)=\op(1)$.

\noindent\textbf{Step 3.} In this and the next steps, we bound $|\hat Q^{-1}-Q^{-1}|$. Note that
\begin{align}
|\hat Q - Q|=&\frac{1}{NM} \sumi\sumj \hat v_{ij}^2  - \Ep [v_{11}^2]
=&
\frac{1}{NM} \sumi\sumj ( v_{ij}^2 - \Ep[v_{11}^2] ) 
+
\frac{1}{NM} \sumi\sumj (\hat v_{ij}^2 - v_{ij}^2 ).\label{eq:var_Q}
\end{align}
The current step bounds the first term on the right-hand side, and Step 4 below bounds the second term on the right-hand side.
With the notation $f(Z_{ij})=v_{ij}^2$, the first term on right-hand size becomes
\begin{align*}
\frac{1}{\sqrt{\C}}\GC f=\frac{1}{NM} \sumi\sumj ( v_{ij}^2 - \Ep[v_{11}^2] ) .
\end{align*}
Applying Lemma \ref{lemma:hajek} under Assumptions \ref{a:sampling} and \ref{a:moments} (1) suggests that its H\'ajek projection equals
\begin{align*}
\frac{1}{\sqrt{\C}}H_n f
&= \frac{1}{N} \sumi \Ep[v^2_{i1}- \Ep v^2_{11}|U_{i0}] +  \frac{1}{M} \sumi \Ep[v^2_{1j}- \Ep v^2_{11}|U_{0j}]
\\
&=\Op\Big(\frac{1}{\sqrt{N}}+\frac{1}{\sqrt{M}}\Big)=\Op\Big(\frac{1}{\sqrt{\C}}\Big),
\end{align*}
where the second equality follows from Lyapunov's CLT applied under Assumption \ref{a:moments} (1) -- note that
Assumption \ref{a:moments} implies the third moments for both terms of the right-hand side to be bounded. 
Assumption \ref{a:moments} and the second claim in Lemma \ref{lemma:hajek} imply that $V(H_n f)=V(\GC f)+O(\C^{-1})$.  
Since $H_n f$ is a projection of $\GC f$, we obtain 
$\frac{1}{\sqrt{\C}}\GC f=\Op(\C^{-1/2})=\op(1)$.

\noindent\textbf{Step 4.} To bound the second term on the RHS of equation (\ref{eq:var_Q}), note that
\begin{align*}
\frac{1}{NM} \sumi\sumj (\hat v_{ij}^2 - v_{ij}^2 )=&
\frac{1}{NM} \sumi\sumj [(X_{ij}\gamma)^2-(X_{ij}\hat \gamma)^2]
+
\frac{2}{NM} \sumi\sumj D_{ij} X_{ij}'(\hat \gamma - \gamma) 
\\
+
\frac{1}{NM} \sumi\sumj (R^D_{ij})^2
&- 
\frac{2}{NM} \sumi\sumj  D_{ij} R_{ij}^D
+
\frac{2}{NM} \sumi\sumj R^D_{ij} X_{ij}'\gamma\\
=&(3a) +(3b) +(3c)+(3d)+(3e).
\end{align*}
The first term can be bounded by
\begin{align*}
|(3a)|=&\Big|\frac{1}{NM} \sumi\sumj [(X_{ij}\gamma)^2-(X_{ij}\hat \gamma)^2]\Big|
\\
\lesssim_\Pr & \sup_{\substack{\|\delta\|=1\\ \|\delta\|_0 \le Cs}} \delta' \Big(\frac{1}{NM}\sumi\sumj X_{ij}X_{ij}'\Big)\delta \cdot \|\hat \gamma - \gamma \|^2
+
2  \sqrt{\frac{1}{NM}\sumi\sumj (X_{ij}'\gamma)^2}
\cdot \|\hat \gamma - \gamma \|\\
\le& \sqrt{\semax{Cs}}\|\hat \gamma-\gamma\|^2 
+
2\Op\Big(  \sqrt{\frac{1}{NM}\sumi\sumj \Ep(X_{ij}'\gamma)^2}\Big) \|\hat\gamma - \gamma\|
\lesssim_\Pr \sqrt{\frac{s \log a}{\C}},
\end{align*}
where the first inequality follows from Assumption \ref{a:sparsity} (1), Lemma \ref{lemma:empirical_pre-sparsity}, and Lemma 3.1 in the supplementary appendix of \cite{vdG14}, and 
the third follows from Theorem \ref{theorem:rates} and Assumption \ref{a:sparse_eigenvalues}. 
An application of Cauchy-Schwartz's inequality and Theorem \ref{theorem:rates} gives
$(3b)=\Op(\sqrt{s\log a/\C})$. 
$(3c)=s/\C$ follows from Assumption \ref{a:sparsity}.
Cauchy-Schwartz's inequality and Assumptions \ref{a:moments}(1), \ref{a:sparsity}(2) lead to
$(3d)\le \|(NM)^{-1/2}R^D\|\sqrt{\Ep[ D_{11}^2]}= \Op(\sqrt{s/\C}) O(1)$. 
Using the property of the projection and a similar argument to that of $(3d)$, we conclude that $(3e)\le \|(NM)^{-1/2}R^D\|\sqrt{\Ep[ D_{11}^2]}= \Op(\sqrt{s/\C}) O(1)$. This along with the conclusion of Step 3 show $|\hat Q - Q|=\op(1)$. Applying the continuous mapping theorem under Assumption \ref{a:moments}(1) then gives $|\hat Q^{-1} -Q^{-1}|=\op(1)$ .

\noindent\textbf{Step 5.} Finally, $|\hat \Gamma|\le |\Gamma| +|\hat \Gamma - \Gamma|=\Op(1)$ following the bounds from Steps 1 and 2 and Assumption \ref{a:moments} (1). Similarly, $|\hat Q^{-1}|\le | Q^{-1}| + |\hat Q^{-1}- Q^{-1}|$ are both bounded following Assumption \ref{a:moments} (1) and Steps 3 and 4.
\end{proof}

\section{Auxiliary Lemmas}\label{sec:auxiliary_lemmas}
The following Lemma is an immediate consequence of Theorem 5.1 of \cite{CCK14} and Lemma 8 of \cite{CCK15}.
\begin{lemma}[A Concentration Inequality]\label{lemma:concentration_inequality_CCK}
Let $(X_i)_{i\in [n]}$ be $p$-dimensional independent random vectors and let $B=\sqrt{E[\max_{i\in [n]}\|X_i\|^2_\infty] }$ and $\sigma^2=\max_{j\in[p]}\frac{1}{n}\sum_{i=1}^n E|X_{ij}|^2$. Then with probability at least $1-C(\log n)^{-1}$,
\begin{align*}
 \max_{j\in [p]}\Big| \frac{1}{n}\sum_{i=1}^n (|X_{ij}| - E|X_{ij}|)\Big| \lesssim \sqrt{\frac{\sigma^2\log (p\vee n)}{n} }
+
\frac{B \log( p\vee n)}{n}.
\end{align*}
\end{lemma} 

The following is an immediate result of Lemma 10 of \cite{BCCH12} with $n=NM$, $\lambda=C NM\sqrt{\log a/\C}$ for $C>1$ and $c_s=\sqrt{s/\C}$.

\begin{lemma}[Sparsity Bound for Lasso]\label{lemma:empirical_pre-sparsity}
Consider lasso estimator (\ref{eq:lasso_main}) and suppose Assumption \ref{a:sparsity} (1)--(2) and \ref{a:sparse_eigenvalues}. Then suppose $\lambda_2/NM\ge c\|(NM)^{-1}\sumi\sumj X_{ij}v_{ij}\|_\infty$ w.p. $1-o(1)$, then denote $\hat s=\supp(\hat \gamma)$, we have $\hat s \lesssim_\Pr s$. Similar result holds for lasso estimator (\ref{eq:lasso_second}) as well.
\end{lemma}

\newpage


\begin{table}
	\centering
		\begin{tabular}{cccccccccccc}
			\hline\hline
			   &    &     && \multicolumn{4}{c}{Statistics}&& \multicolumn{3}{c}{95\% Coverage}\\
		  \cline{5-8}\cline{10-12}
			$N$& $M$& Dim && Avg   & Bias  & SD    & RMSE  && 0-Way & 1-Way & 2-Way\\
			\hline
			10 & 10 & 100 && 0.491 &-0.009 & 0.172 & 0.172 && 0.808 & 0.797 & 0.919\\
			20 & 20 & 100 && 0.499 &-0.001 & 0.076 & 0.076 && 0.855 & 0.858 & 0.964\\
			40 & 40 & 100 && 0.501 & 0.001 & 0.045 & 0.045 && 0.792 & 0.848 & 0.959\\
			\\
			10 & 10 & 200 && 0.480 &-0.020 & 0.356 & 0.357 && 0.753 & 0.747 & 0.877\\
			20 & 20 & 200 && 0.498 &-0.002 & 0.075 & 0.075 && 0.863 & 0.860 & 0.962\\
			40 & 40 & 200 && 0.500 & 0.000 & 0.041 & 0.041 && 0.830 & 0.859 & 0.962\\
			\\					
			10 & 10 & 400 && 0.459 &-0.041 & 0.744 & 0.745 && 0.682 & 0.690 & 0.822\\
			20 & 20 & 400 && 0.496 &-0.004 & 0.079 & 0.079 && 0.846 & 0.841 & 0.951\\
			40 & 40 & 400 && 0.500 & 0.000 & 0.037 & 0.037 && 0.869 & 0.876 & 0.970\\
			\\						
			10 & 10 & 800 && 0.438 &-0.062 & 0.959 & 0.961 && 0.615 & 0.634 & 0.764\\
			20 & 20 & 800 && 0.492 &-0.008 & 0.086 & 0.086 && 0.808 & 0.803 & 0.930\\
			40 & 40 & 800 && 0.499 &-0.001 & 0.037 & 0.037 && 0.870 & 0.871 & 0.968\\
			\\						
			10 & 10 &1600 && 0.394 &-0.106 & 1.140 & 1.140 && 0.555 & 0.585 & 0.704\\
			20 & 20 &1600 && 0.487 &-0.013 & 0.098 & 0.099 && 0.763 & 0.762 & 0.903\\
			40 & 40 &1600 && 0.498 &-0.002 & 0.038 & 0.038 && 0.862 & 0.859 & 0.964\\
			\hline\hline
			\\
		\end{tabular}
	\caption{
	Simulation results. The first three columns indicate the two-way sample sizes $(N,M)$ and the dimension (Dim) of $(\alpha,\beta')'$. The next four columns report simulation statistics for $\widetilde\alpha$, including the average (Avg), bias (Bias), standard deviation (SD), and root mean square error (RMSE). The last three columns report 95\% coverage frequencies of $\alpha$ with the heteroskedasticity robust variance estimator (0-Way), the one-way cluster-robust variance estimator (1-Way), and our multi-way cluster-robust variance estimator (2-Way). The data generating parameters are set to 
$(\omega^{x}_1, \omega^{x}_2) = (0.25,0.25)$,
$(\omega^{\varepsilon}_1,\omega^{\varepsilon}_2) = (0.25,0.25)$, and
$\rho = 0.50$.
The results are based on 25,000 Monte Carlo iterations for each row.}
	\label{tab:simulation_results}
\end{table}

\begin{table}
	\centering
		\begin{tabular}{cccccccc}
		\hline\hline
		  \multicolumn{2}{c}{Variables}&Number of&\multicolumn{2}{c}{Cluster Size}&Original&Lasso\\
		\cline{1-2}\cline{4-5}
			$Y$      & $D$         &Observations&$N$&$M$& Estimates  & Estimates\\
		\hline
			Trust of & Slave       & 20,027 & 185 & 1,257 & -0.00068 & -0.00083\\
			Neighbors& Exports     &        &     &       & (0.00015)& (0.00022)\\
		\hline
			Trust of & Exports/    & 20,027 & 185 & 1,257 & -0.019 & -0.025\\
			Neighbors& Area        &        &     &       & (0.005)& (0.007)\\
		\hline
			Trust of & Exports/    & 17,644 & 157 & 1,214 & -0.531 & -0.684\\
			Neighbors& Population  &        &     &       & (0.147)& (0.232)\\
		\hline
			Trust of & Log Slave   & 20,027 & 185 & 1,257 & -0.037 & -0.045\\
			Neighbors& Exports     &        &     &       & (0.014)& (0.021)\\
		\hline
			Trust of & Log Exports/& 20,027 & 185 & 1,257 & -0.159 & -0.210\\
			Neighbors& Area        &        &     &       & (0.034)& (0.050)\\
		\hline
			Trust of & Log Exports/& 17,644 & 157 & 1,214 & -0.743 & -0.957\\
			Neighbors& Population  &        &     &       & (0.187)& (0.304)\\
		\hline\hline
		\\
		\end{tabular}
	\caption{Estimates of the effects of slave trade on mistrust in Africa. The first two columns indicate which measures of the dependent and explanatory variables are used. The next three columns show the number of observations, the number of ethnic groups ($N$), and the number of districts ($M$). The last two columns show the original estimates obtained under the prototype model by \citet[][Table 1]{NunnWantchekon11} and corresponding lasso estimates obtained under more flexible model specification by our method.}
	\label{tab:results_nunn}
\end{table}

\begin{table}
	\centering
		\begin{tabular}{ccccccccc}
		\hline\hline
		  \multicolumn{2}{c}{Variables}&No.&\multicolumn{2}{c}{Cluster Size}&Population&Original&Lasso\\
		\cline{1-2}\cline{4-5}
			$Y$      & $D$         &Obs.&$N$&$M$&Density& Estimates  & Estimates\\
		\hline
			Light  & Jurisdictional& 682 & 93 & 48 &No & 0.2794 & 0.2266\\
			Density& Hierarchy     &     &    &    &   &(0.0852)&(0.0797)\\
		\hline
			Light  & Jurisdictional& 682 & 93 & 48 &Yes& 0.1766 & 0.1649\\
			Density& Hierarchy     &     &    &    &   &(0.0501)&(0.0541)\\
		\hline
			Light  & Political     & 682 & 93 & 48 &No & 0.5049 & 0.4158\\
			Density& Centralization&     &    &    &   &(0.1573)&(0.1451)\\
		\hline
			Light  & Political     & 682 & 93 & 48 &Yes& 0.3086 & 0.2985\\
			Density& Centrilization&     &    &    &   &(0.0972)&(0.1080)\\
		\hline\hline
		\\
		\end{tabular}
	\caption{Estimates of the effects of pre-colonial institutions on regional development in Africa. The first two columns indicate which measures of the dependent and explanatory variables are used. The next three columns show the number of observations, the number of ethnic groups ($N$), and the number of districts ($M$). The next column indicates a control for population density. The last two columns show the original estimates obtained under the prototype model by \citet[][Table 3]{Michalopoulos_Papaioannou2013} and corresponding lasso estimates obtained under more flexible model specification by our method.}
	\label{tab:results_michalopoulos}
\end{table}

\end{document}